# Uncovering Collective Modes Underlying the Giant Dielectric Response of Ferroelectric Nematic Liquid Crystals

Kazuma Nakajima[1*], Hirokazu Kamifuji[1], Masanori Ozaki[1], Hirotsugu Kikuchi[2], and Kenjiro Fukuda[1*]

[1]Division of Electrical, Electronic and Infocommunications Engineering, Graduate School of Engineering, The University of Osaka, 2-1 Yamadaoka, Suita, Osaka 565-0871, Japan

[2]Institute for Materials Chemistry and Engineering, Kyushu University, Kasuga, Fukuoka 816-8580, Japan

*nakajima.kazuma@eei.eng.osaka-u.ac.jp, fukuda@eei.eng.osaka-u.ac.jp

**Abstract**

Ferroelectric nematic liquid crystals (FNLCs) are polar fluids in which spontaneous polarization coexists with nematic orientational order, giving rise to unusual dielectric and electromechanical responses. However, the collective modes underlying their giant dielectric response remain unclear. Here, we show that this response originates from the superposition of two distinct relaxation modes rather than a single process. Dielectric spectroscopy reveals that the low-frequency mode exhibits soft-mode-like behavior associated with short-axis molecular rotation, whereas the high-frequency mode corresponds to a Goldstone-like phase displacement of an effective transverse polarization component rotating around the director. These assignments are supported by systematic analyses of temperature, electric-field, cell-thickness, and alignment-layer dependences. Our results demonstrate that the giant dielectric response of ferroelectric nematics reflects multiple collective polarization dynamics with different symmetries and restoring forces, providing a framework for interpreting dielectric spectra in polar nematic fluids.

## 1. Introduction

Ferroelectric nematic liquid crystals (FNLCs) have recently been discovered as a new liquid-crystalline phase in which spontaneous polarization coexists with nematic orientational order, and have had a major impact on the fields of liquid-crystal physics and soft-matter physics [1–5]. Unlike conventional nematic liquid crystals, FNLCs possess macroscopic spontaneous polarization and exhibit characteristic physical properties that have not been observed in ordinary liquid crystals, such as piezoelectric/inverse piezoelectric effects[6–8], pyroelectric responses[9,10], and strong nonlinear optical effects[11–15]. These unusual properties are considered to reflect the essential nature of FNLCs as a new class of polar fluids combining the characteristics of ferroelectrics and fluids. Therefore, it is extremely important to understand the fundamental physics behind them.

In ordered phases with spontaneous symmetry breaking, it is generally crucial to understand not only the static ordered structure but also the collective motions of the degrees of freedom responsible for the order. For example, spin waves in magnetic systems and soft modes in ferroelectrics have played central roles in understanding the formation mechanisms of ordered states and the nature of phase transitions. Similarly, in liquid crystals, elastic deformations associated with director orientational order and fluctuations of order parameters are fundamental degrees of freedom governing phase transitions, electro-optic responses, and viscoelastic responses. Therefore, to understand the physical properties of FNLCs, it is necessary to clarify what types of collective modes exist in the polar ordered state where spontaneous polarization and the director are coupled, and how these modes respond to external fields and boundary conditions.

In liquid-crystal physics, Goldstone modes and soft modes are particularly important concepts for understanding ferroelectric order[16–20]. In ferroelectric smectic C* liquid crystals (SmC* FLC), molecular tilt with respect to the layer normal forms a conically degenerate orientational state. Azimuthal fluctuations along this degenerate free-energy surface appear as Goldstone modes arising from continuous symmetry breaking. In contrast, fluctuations in the tilt angle or in the magnitude of the ferroelectric order parameter near the phase transition are observed as soft modes. These two modes are clearly distinguished in dielectric spectroscopy and have played central roles in understanding phase transitions and electro-optic responses in ferroelectric liquid crystals.

The terms Goldstone-like mode and soft-mode-like mode have also been used for FNLCs. FNLCs are polar nematic phases without a layered structure[1,21], and their order-parameter space differs from the conical degeneracy in SmC* FLCs. Therefore, dielectric relaxation modes observed in FNLCs cannot be simply mapped onto the modes known in SmC* FLCs. In particular, the fact that a mode appears at low frequency or exhibits a large dielectric strength is not sufficient evidence to identify it as a Goldstone-like mode. To discuss a Goldstone-like mode in FNLCs, it is necessary to demonstrate that the mode corresponds to a collective fluctuation along a degenerate orientational degree of freedom of the polar director.

Indeed, recent theoretical and experimental studies have suggested the existence of collective motions characteristic of FNLC materials. Dynamic light scattering studies have reported that the fluctuation spectrum changes significantly across the transition from the smectic $Z_A$ ($SmZ_A$) phase to the ferroelectric nematic ($N_F$) phase[22]. In addition, it has been suggested that spontaneous splay and bend deformations in FNLCs are strongly coupled to polarization, giving rise to unusual collective modes that do not exist in conventional nematic liquid crystals[22–24]. These studies suggest that collective excitations in FNLCs are not simple director fluctuations of ordinary nematics, but rather complex motions involving coupled polarization, elasticity, and boundary conditions.

The dielectric response of FNLCs provides one of the most sensitive probes of collective polar dynamics, because it directly reflects fluctuations of spontaneous polarization and molecular orientation. Accordingly, many studies have investigated the dielectric properties of FNLCs using dielectric spectroscopy. In particular, FNLCs exhibit giant dielectric permittivity far exceeding that of conventional nematic liquid crystals, and the origin of this giant dielectric response has been actively debated. The proposed interpretations can be broadly classified into two categories. One is the picture in which giant dielectric permittivity arises from the transport of polarization charges and their accumulation at interfaces, represented by the polarization-external capacitance Goldstone (PCG) model[25–28]. In this model, the observed giant dielectric response is understood primarily as an interfacial capacitance or Maxwell–Wagner-type effect. The other is the high-$\varepsilon$ model, in which FNLCs are regarded as polar fluids possessing intrinsically giant dielectric permittivity, and the dielectric response originates from collective fluctuations of spontaneous polarization[29,30]. Regardless of whether the giant response is interpreted mainly as an interfacial polarization-charge effect or as an intrinsic high-permittivity response, the dielectric spectra themselves must be decomposed into physically meaningful relaxation modes.

However, the collective modes and their molecular motions underlying this giant dielectric response remain insufficiently understood. One source of this ambiguity is that the dielectric response in the frequency range where giant permittivity appears has often been analyzed as a single relaxation process. Previous studies that considered more than one dielectric response in FNLC spectra mainly reported additional modes with relaxation frequencies around 10 Hz[31] or around $10^7$ Hz[32], which lie outside the frequency range of the giant dielectric response. Thus, the possible coexistence of multiple collective relaxation modes within the giant dielectric response of FNLCs has not yet been systematically examined. If such modes overlap, a single-mode analysis cannot correctly identify the physical origin of each mode and may lead to an incorrect assignment of relaxation modes.

Here, we demonstrate that the giant dielectric response of a dioxane-based FNLC mixture (DIO mixture) originates not from a single relaxation process, but from the superposition of two distinct collective modes. More specifically, we reveal that the low-frequency mode is a soft-mode-like fluctuation associated with short-axis molecular rotation, whereas the high-frequency mode represents a Goldstone-like phase displacement of an effective transverse polarization component rotating around the director. These microscopic origins are established by systematically analyzing the dielectric spectra as functions of temperature near the phase transition, electric field, cell thickness, and alignment conditions. Consequently, this work provides a framework for reinterpreting the giant dielectric response of FNLCs not as a single relaxation phenomenon, but as a superposition of multiple collective modes with different symmetries and restoring forces.

## 2. Results

### 2.1 Identification of two relaxation modes in FNLC

To clarify the relaxation modes responsible for the dielectric response of FNLCs, we analyzed the complex permittivity spectra using Cole–Cole-type relaxation functions. The cells used in this study consisted of orthogonally crossed ITO electrodes with a width of 2 mm, resulting in an effective electrode area of 4 mm$^2$. The complex permittivity was described as a superposition of Cole–Cole relaxation processes as follows:

$$\varepsilon^*(\omega) = \varepsilon_\infty + \sum_{i=1}^{N} \frac{\Delta\varepsilon_{s-\infty,i}}{1 + (i\omega\tau_i)^{1-\alpha_i}} - \frac{i\sigma_{\mathrm{DC}}}{\varepsilon_0\omega}, \tag{1}$$

where $\varepsilon_\infty$ is the high-frequency permittivity, and $\Delta\varepsilon_{s-\infty,i}$, $\tau_i$, and $\alpha_i$ are the dielectric strength, relaxation time, and Cole–Cole exponent of the $i$-th relaxation mode, respectively. Here, $\sigma_{DC}$ is the dc conductivity and $\varepsilon_0$ is the vacuum permittivity. In the present analysis, $\varepsilon_\infty$ was fixed at 10, and $\sigma_{DC}$ was fixed to the value estimated from the low-frequency region. The parameters $\Delta\varepsilon_{s-\infty,i}$, $\tau_i$, and $\alpha_i$ were optimized for each relaxation mode by Cole–Cole fitting. The corresponding Cole–Cole plots are shown in Fig. S1.

Figures 1(a–c) show representative dielectric spectra measured at $75\,°\mathrm{C}$ and the corresponding fitting results for a 9.05-µm-thick cell without an alignment layer, a 9.51-µm-thick cell coated with LIA-03, which is commonly used as an azobenzene-based photoalignment material, and an 8.90-µm-thick cell coated with an unrubbed polyimide layer, AL1254, commonly used for planar alignment, respectively. In all cells, no rubbing treatment or photoalignment was applied, and a FNLC mixture consisting of three fluoro-4-(3,4,5-trifluorophenyl)phenyl 2,6-difluoro-4-(trans-5-n-propyl-1,3-dioxane-2-yl)benzoate derivatives with different alkyl-chain lengths (DIO mixture) was used.

It should first be noted that the magnitude of the low-frequency real permittivity strongly depends on the alignment-layer condition. The low-frequency value of $\varepsilon'$ is approximately $6.2 \times 10^3$ in the cell without an alignment layer, whereas it increases to approximately $1.4 \times 10^4$ in the LIA-03-coated cell and decreases to approximately $5.8 \times 10^2$ in the AL1254-coated cell. Thus, the alignment layer does not simply shift the overall magnitude of the dielectric response, but drastically changes the low-frequency dielectric response in an alignment-layer-dependent manner. This pronounced variation indicates that the observed giant dielectric response is strongly influenced by interfacial conditions and cannot be regarded as an alignment-independent dielectric response.

Because dielectric responses in FNLCs have often been analyzed using a single relaxation process, we then attempted to fit the spectra using a single-relaxation model, hereafter referred to as the one-arc model. However, as shown in Figs. 1(a–c), the one-arc model fails to reproduce both the transition region of the real permittivity $\varepsilon'$ and the peak shape of the imaginary permittivity $\varepsilon''$. In particular, the broad frequency dispersion of $\varepsilon''$ cannot be described by a single Cole–Cole relaxation, indicating that a single-mode picture is insufficient to account for the experimental spectra. In contrast, the two-arc model, which assumes two independent relaxation processes, reproduces the frequency dependences of both $\varepsilon'$ and $\varepsilon''$ over the entire measured frequency range. This result suggests that the dielectric response of the DIO mixture consists of at least two relaxation modes. Importantly, the need for the two-arc model is observed for all alignment-layer conditions, despite the large difference in the magnitude of the low-frequency permittivity.

To quantitatively evaluate this result, we compared the one-arc and two-arc models using the Bayesian information criterion (BIC). The BIC evaluates the balance between goodness of fit and model complexity; a lower BIC indicates a statistically more favorable model, and a difference of $\Delta\mathrm{BIC} > 10$ is generally regarded as strong evidence for the model with the lower BIC. Here, we define $\Delta\mathrm{BIC} = \mathrm{BIC}_{1\mathrm{arc}} - \mathrm{BIC}_{2\mathrm{arc}}$, so that $\Delta\mathrm{BIC} > 10$ indicates that the two-arc model is favored. Figures 1(d–f) show the temperature dependence of $\Delta\mathrm{BIC}$ for the 9.05-μm-thick cell without an alignment layer, the 9.51-μm-thick LIA-03-coated cell, and the 8.90-μm-thick AL1254-coated cell, respectively. For all alignment conditions and measured temperatures, $\Delta\mathrm{BIC}$ was positive. Moreover, $\Delta\mathrm{BIC}$ exceeded approximately 90 over the entire temperature range and was much larger than 100 over a wide temperature range. Around the $\mathrm{SmZ_A}$–$\mathrm{N_F}$ phase transition near 84 °C, $\Delta\mathrm{BIC}$ exhibited a clear change: it increased across the transition in the cell without an alignment layer, whereas it decreased in the LIA-03- and AL1254-coated cells. Nevertheless, even in the $\mathrm{N_F}$ phase, $\Delta\mathrm{BIC}$ remained comparable to that in the $\mathrm{SmZ_A}$ phase, where the spectra have been regarded as requiring a two-arc description. These results show that the giant dielectric response in the $\mathrm{N_F}$ phase cannot be described by a single relaxation process, but is more appropriately understood as the superposition of at least two distinct relaxation modes.

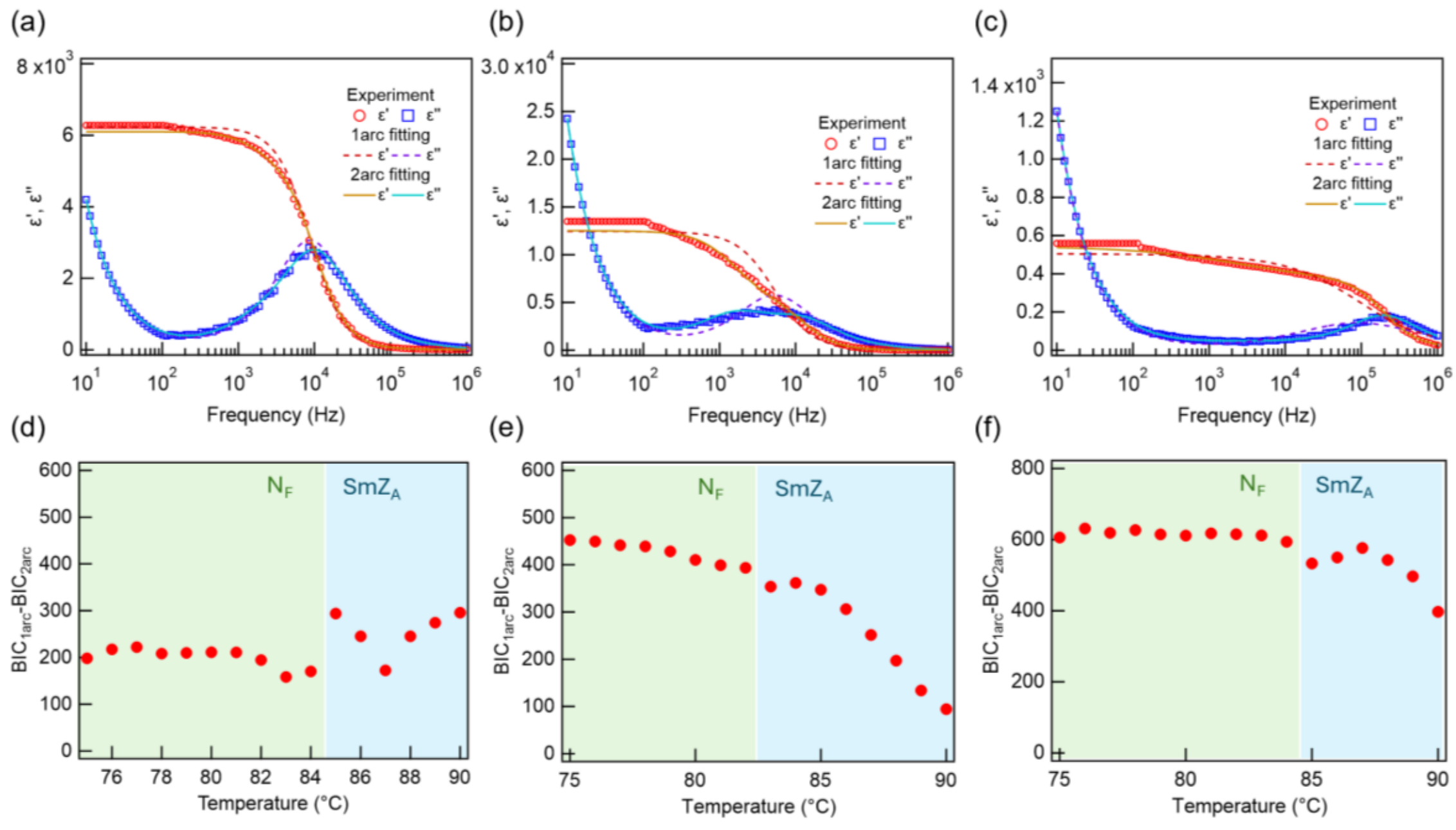

**Figure 1** Identification of two dielectric relaxation modes in the DIO mixture. (a–c) Representative dielectric spectra measured at 75 °C for (a) a 9.05-μm-thick cell without an alignment layer, (b) a 9.51-μm-thick LIA-03-coated cell, and (c) an 8.90-μm-thick AL1254-coated cell. Open symbols represent the experimental values of the real ($\varepsilon'$) and imaginary ($\varepsilon''$) parts of the complex dielectric permittivity. Solid lines show fitting results obtained using the conventional one-relaxation (one-arc) model and the two-relaxation (two-arc) model. (d–f) Temperature dependence of the Bayesian information criterion difference, $\Delta\mathrm{BIC} = \mathrm{BIC}_{1\mathrm{arc}} - \mathrm{BIC}_{2\mathrm{arc}}$, for the cells (d) without an alignment layer, (e) with LIA-03 , and (f) with AL1254. Positive $\Delta\mathrm{BIC}$ values indicate statistical preference for the two-arc model.

**2.2 Temperature dependence of the two relaxation modes**

Figure 2 shows the temperature dependences of the dielectric strength ($\Delta\varepsilon_{s-\infty}$) and relaxation frequency ($f_r$) obtained from the one-arc and two-arc models for a 9.05-μm-thick cell without an alignment layer, a 9.51-μm-thick cell coated with LIA-03, and an 8.90-μm-thick cell coated with AL1254. Here, $m_H$ and $m_L$ denote the high- and low-frequency relaxation modes, respectively. Since two relaxation modes were clearly resolved in the $SmZ_A$ phase, only the two-arc model was applied in this phase. The measurements were performed during cooling at a rate of 0.2 °C $min^{-1}$, and the corresponding dielectric spectra are presented in Fig. S2.

We first consider the cell without an alignment layer (Figs. 2(a,d)). Within the conventional single-relaxation picture, the dielectric strength has been interpreted as increasing continuously from the $SmZ_A$ phase into the $N_F$ phase, leading to the emergence of the giant dielectric response. Indeed, the dielectric strength obtained from the one-arc model exhibits nearly identical values at 85 °C in the $SmZ_A$ phase and 84 °C in the $N_F$ phase, giving the apparent impression of a continuous evolution across the phase transition. However, the slope of the dielectric strength near 85 °C in the $SmZ_A$ phase is markedly different from that near 84 °C in the $N_F$ phase. Moreover, the relaxation frequency increases by approximately one order of magnitude, from $\sim 10^3$ Hz in the $SmZ_A$ phase to $\sim 10^4$ Hz in the $N_F$ phase. These observations make it difficult to interpret the low-frequency relaxation mode in the $SmZ_A$ phase as continuously transforming into a single relaxation process in the $N_F$ phase, or to regard the apparent continuity of the dielectric strength as evidence of a continuous phase-transition-related response.

The two-arc analysis resolves the origin of this apparent continuity. The dielectric strength of the low-frequency mode ($\Delta\varepsilon_{s-\infty,L}$) reaches a maximum near 85 °C and subsequently decreases upon entering the $N_F$ phase (Fig. 2(a)). In contrast, the dielectric strength of the high-frequency mode ($\Delta\varepsilon_{s-\infty,H}$) increases abruptly near 84 °C. Consequently, the decrease in $\Delta\varepsilon_{s-\infty,L}$ and the increase in $\Delta\varepsilon_{s-\infty,H}$ compensate each other, causing the one-arc model to yield similar dielectric strengths at 85 °C and 84 °C. Therefore, the dielectric anomaly previously interpreted as a continuous single-mode response is, in fact, a superposition of two relaxation modes with distinct temperature dependences. The relaxation frequencies exhibit similarly contrasting behavior. The relaxation frequency of the low-frequency mode ($f_{r,L}$) shows a minimum near 84 °C (Fig. 2(d)). In contrast, the relaxation frequency of the high-frequency mode ($f_{r,H}$) decreases upon cooling in the $SmZ_A$ phase, undergoes a discontinuous drop near 84 °C, and subsequently increases again in the $N_F$ phase. Such behavior cannot be extracted from the one-arc model and becomes evident only after separating the two relaxation modes.

The results for the LIA-03-coated cell are shown in Figs. 2(b,e). The most notable difference from the cell without an alignment layer is that the low-frequency mode remains the dominant dielectric contribution even in the $N_F$ phase. Specifically, $\Delta\varepsilon_{s-\infty,L}$ increases continuously from the $SmZ_A$ phase and retains a large value throughout the $N_F$ phase. Although $\Delta\varepsilon_{s-\infty,H}$ also increases across the phase transition, it remains smaller than that of the low-frequency mode (Fig. 2(b)). The temperature dependence of the relaxation frequencies is qualitatively similar to that observed in the cell without an alignment layer: ($f_{r,L}$) exhibits a minimum near the transition temperature, whereas ($f_{r,H}$) increases with decreasing temperature in the $N_F$ phase (Fig. 2(e)).

The results for the AL1254-coated cell are shown in Figs. 2(c,f). The temperature dependence of the dielectric strengths is qualitatively similar to that observed in the cell without an alignment layer, with an increase in the high-frequency mode upon entering the $N_F$ phase. However, $\Delta\varepsilon_{s-\infty,H}$ remains remarkably small, reaching only approximately 350 even in the $N_F$ phase (Fig. 2(c)). In contrast, its relaxation frequency remains as high as $\sim 2 \times 10^5$ Hz, which is comparable to that observed for the high-frequency mode in the $SmZ_A$ phase (Fig. 2(f)). Thus, in the AL1254-coated cell, the amplitude of the high-frequency mode is strongly suppressed, whereas its dynamics are substantially accelerated. This finding indicates that the high-frequency mode is strongly influenced by interfacial conditions and surface anchoring. In contrast, $\Delta\varepsilon_{s-\infty,L}$ is also reduced compared with the cells without an alignment layer and with LIA-03, indicating that its amplitude is affected by the alignment condition. However, its relaxation frequency remains nearly identical to those observed under the other alignment conditions. This result indicates that, unlike the dielectric strength, the dynamics of the low-frequency mode are largely insensitive to the alignment layer.

These results demonstrate that the two relaxation modes respond fundamentally differently to the phase transition. The low-frequency mode evolves continuously across the $SmZ_A$–$N_F$ transition, exhibiting a characteristic maximum in dielectric strength and a minimum in relaxation frequency near the transition temperature. In contrast, the high-frequency mode exhibits discontinuous changes in both dielectric strength and relaxation frequency upon entering the $N_F$ phase. Such contrasting temperature dependences indicate that the two modes do not originate from a common relaxation process but instead represent independent collective modes with distinct physical origins, symmetries, and restoring mechanisms.

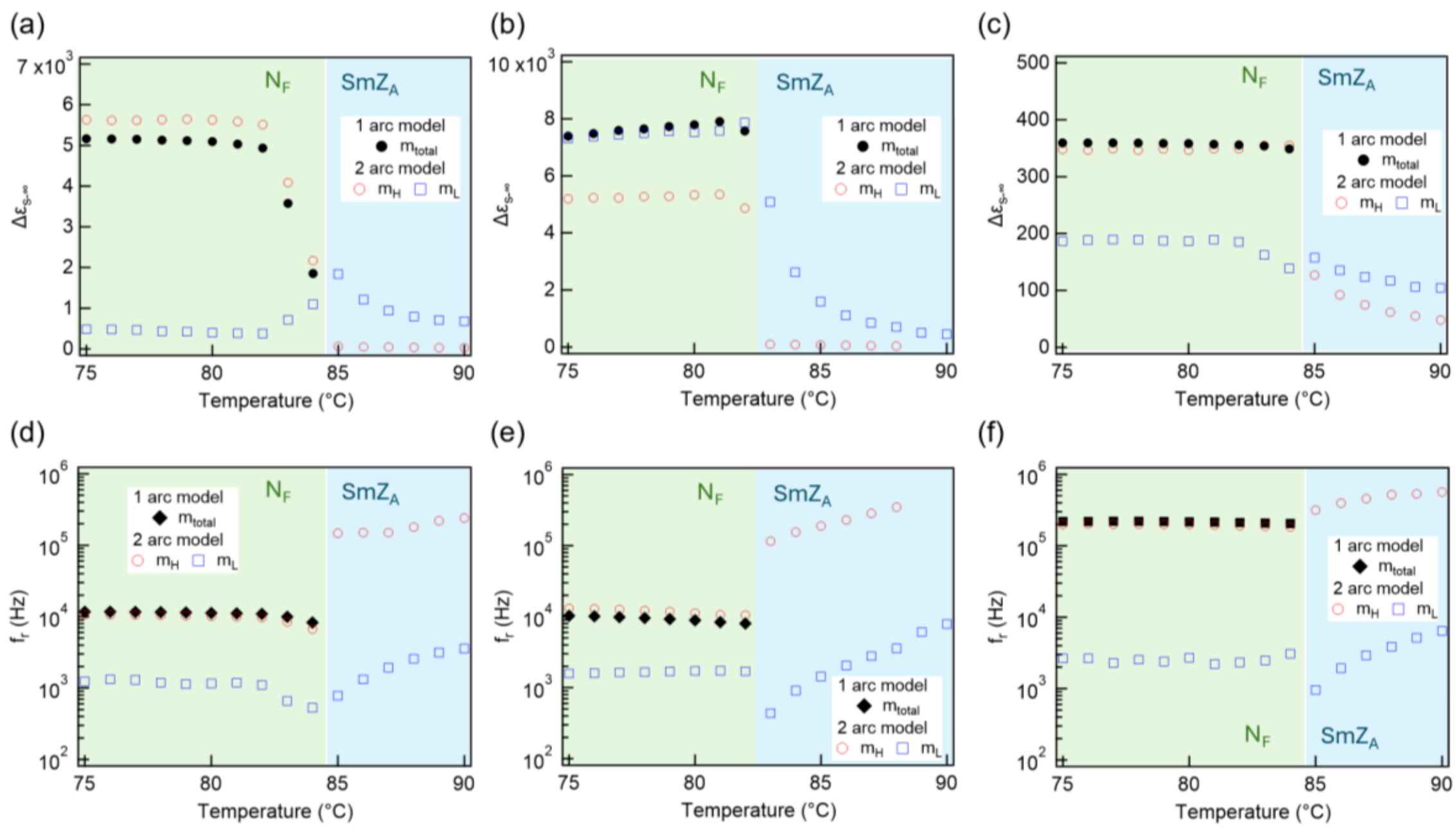

**Figure 2** Temperature dependence of the two relaxation modes. (a–c) Temperature dependence of the dielectric strength ($\Delta\varepsilon_{s-\infty}$) obtained from one-arc and two-arc analyses for cells (a) a 9.05-µm-thick cell without an alignment layer, (b) a 9.51-µm-thick LIA-03-coated cell, and (c) an 8.90-µm-thick AL1254-coated cell. Black symbols denote the dielectric strength obtained from the one-arc model, while red and blue symbols correspond to the high-frequency ($m_H$) and low-frequency ($m_L$) modes extracted from the two-arc model, respectively. (d–f) Corresponding relaxation frequencies ($f_r$) of the modes shown in (a–c). The low-frequency mode evolves continuously across the SmZ$_A$–N$_F$ phase transition, whereas the high-frequency mode exhibits discontinuous changes in both dielectric strength and relaxation frequency.

### 2.3 Electric-field dependence of the relaxation modes

To clarify the physical origins of the two relaxation modes, dielectric spectroscopy was performed under applied dc voltages. Figures 3(a–c) show the dielectric spectra and fitting results measured at 80 °C for a 14.25-μm-thick cell without an alignment layer under applied voltages of 0, 2.0, and 4.0 V, respectively. To minimize effects associated with voltage-induced ion adsorption at the interfaces, voltage sweeps from 0 to 5.0 V were repeated eight times at 80 °C, at which the sample was in the $N_F$ phase before this experiment. The subsequent measurements were performed after confirming that both the dielectric strength and relaxation frequency had converged (Fig. S3).

In the absence of an applied voltage, the high-frequency mode dominates the dielectric response, and the peak in the imaginary permittivity $\varepsilon''$ appears mainly near the relaxation frequency of this mode. As the voltage increases to 2.0 V, the contribution of the high-frequency mode gradually decreases, and the spectrum becomes a superposition of the high- and low-frequency modes with comparable weights. At 4.0 V, the dielectric strength of the high-frequency mode is strongly suppressed, and the $\varepsilon''$ peak is dominated primarily by the low-frequency mode.

To quantify this field-induced change, the voltage dependences of the dielectric strength and relaxation frequency of each mode were evaluated, as shown in Figs. 3(d,e). These measurements were performed in the $N_F$ phase over the temperature range from 65 to 80 °C; the raw dielectric spectra at 80 °C are shown in Fig. S4. At all measured temperatures, the dielectric strength of the high-frequency mode, $\Delta\varepsilon_{s-\infty,H}$, decreases with increasing voltage and shows a particularly steep reduction above approximately 2.0 V. In contrast, the dielectric strength of the low-frequency mode, $\Delta\varepsilon_{s-\infty,L}$, increases around the same voltage range and becomes larger than $\Delta\varepsilon_{s-\infty,H}$ above approximately 2.5 V. Thus, the dominant dielectric contribution switches from the high-frequency mode to the low-frequency mode under applied voltage. The relaxation frequency of the high-frequency mode, $f_{r,H}$, increases monotonically with increasing voltage. By contrast, the relaxation frequency of the low-frequency mode, $f_{r,L}$, exhibits a nonmonotonic voltage dependence: it increases up to approximately 2.5 V, where the dielectric strength of the low-frequency mode becomes pronounced, and then gradually decreases at higher voltages. These results indicate that the dc electric field modifies not only the amplitude but also the characteristic relaxation dynamics of the two collective modes in different ways.

Figure 3(f) shows the temperature dependence of the dielectric strengths of the high- and low-frequency modes under different applied voltages, measured using a 14.31-μm-thick cell without an alignment layer. The corresponding temperature dependence of the relaxation frequency is shown in Fig. S5. In this measurement, voltage preconditioning was not performed in order to keep the initial condition consistent with the zero-field measurements shown in Fig. 2 and to directly compare the intrinsic relaxation modes. The high-frequency mode is strongly suppressed as the applied voltage increases, whereas the low-frequency mode retains a relatively large dielectric strength. As a result, a dielectric peak that is obscured by the high-frequency contribution under zero field becomes clearly visible under higher voltages. Moreover, the peak temperature shifts to higher temperature with increasing voltage.

This behavior is qualitatively similar to that observed in SmC* FLCs, where the Goldstone mode is suppressed by an applied electric field and the soft mode becomes apparent[20,33–35]. In FNLCs, it has also been reported that a dielectric peak appears near the phase transition when the director is aligned along the cell normal under a dc electric field[36,37]. This behavior is consistent with the soft-mode-like character of the low-frequency mode, as discussed below.

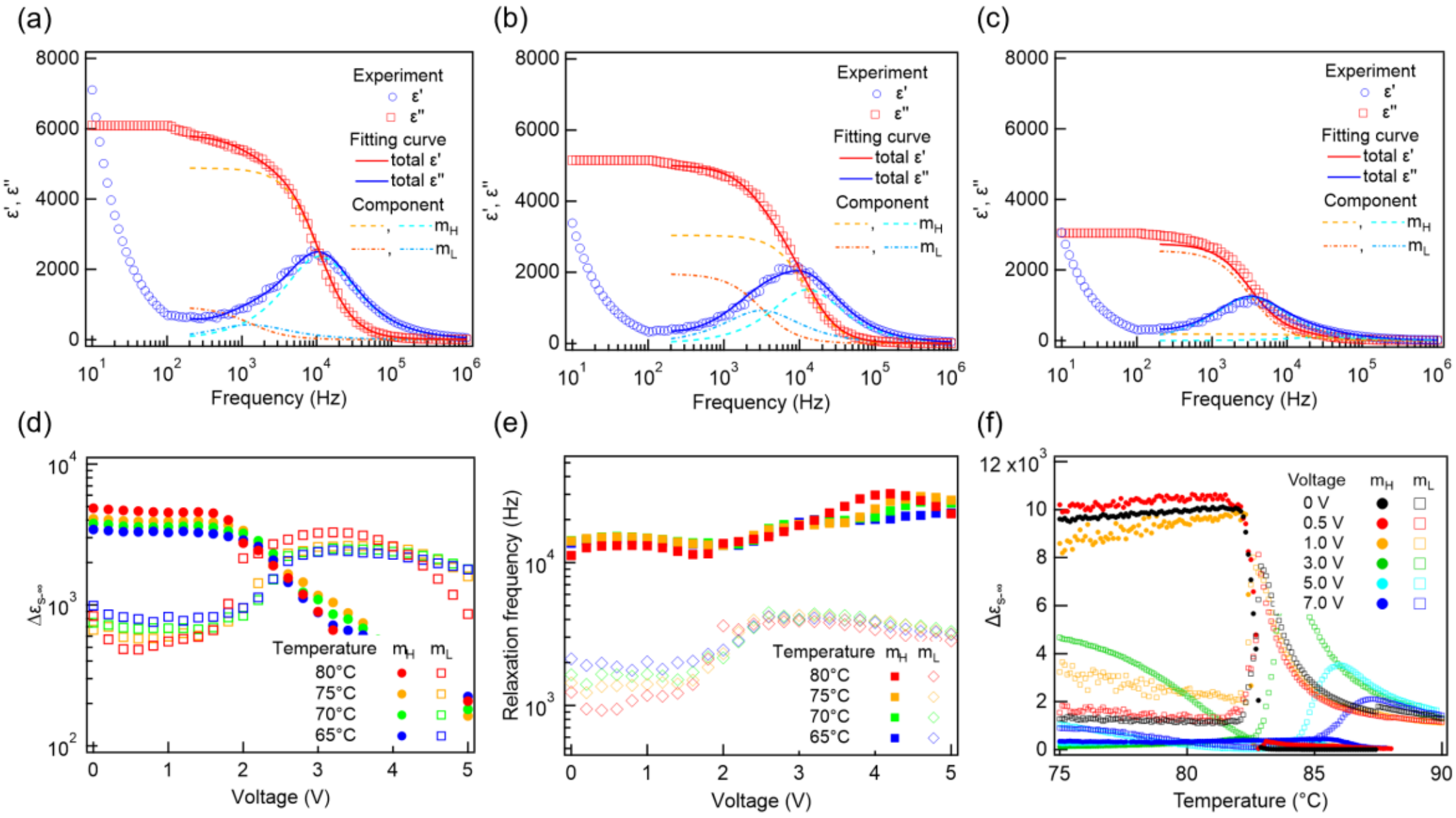

**Figure 3** Electric-field dependence of the dielectric relaxation modes. (a–c) Dielectric spectra measured at 80 °C in a 14.25-μm-thick cell without an alignment layer under dc voltages of (a) 0 V, (b) 2.0 V, and (c) 4.0 V. Solid lines represent the total two-arc fit, while dashed lines show the contributions of the high-frequency ($m_H$) and low-frequency ($m_L$) modes. With increasing dc voltage, the high-frequency contribution is progressively suppressed and the dielectric-loss peak becomes dominated by the low-frequency mode. (d,e) Voltage dependence of (d) dielectric strength and (e) relaxation frequency for the two modes measured at different temperatures in the $N_F$ phase. The dielectric strength of the high-frequency mode decreases markedly above approximately 2 V, whereas the low-frequency mode becomes dominant. (f) Temperature dependence of the dielectric strengths of the two modes under different dc voltages. The high-frequency mode is strongly suppressed by the electric field, while the low-frequency mode retains a large dielectric strength and develops a pronounced dielectric peak near the phase transition. The peak temperature shifts to higher temperature with increasing voltage.

### 2.4 Thickness and alignment dependence of the relaxation modes

To elucidate the origin of the restoring forces associated with the two relaxation modes, dielectric spectroscopy measurements were performed using cells with different thicknesses and alignment conditions. Figures 4(a–f) show the thickness dependences of the dielectric strength ($\Delta\varepsilon_{s-\infty}$) and relaxation frequency ($f_r$) for the high-frequency ($m_H$) and low-frequency modes ($m_L$). Figures 4(a,d) correspond to cells without an alignment layer, Figs. 4(b,e) to LIA-03-coated cells, and Figs. 4(c,f) to AL1254-coated cells.

We first focus on the high-frequency mode (Figs. 4(a–c)). Under all alignment conditions, the dielectric strength $\Delta\varepsilon_{s-\infty,H}$ increases approximately linearly with cell thickness. Furthermore, the inverse of the relaxation frequency, $1/f_{r,H}$, exhibits a nearly linear dependence on thickness, indicating that $f_{r,H}$ is inversely proportional to the cell thickness. These results indicate that the dielectric strength and relaxation frequency of the high-frequency mode are not intrinsic material properties but arise from collective motion.

Furthermore, the product $\Delta\varepsilon_{s-\infty,H}f_{r,H}$ remains nearly constant regardless of cell thickness or alignment condition (Fig. 4(g)). This indicates that the increase in dielectric strength and the decrease in relaxation frequency compensate each other. Therefore, the high-frequency mode follows a common scaling relation governed by the cell thickness, regardless of the alignment layer material.

In contrast, the low-frequency mode exhibits a different behavior. The dielectric strength $\Delta\varepsilon_{s-\infty,L}$ increases approximately linearly with cell thickness under all alignment conditions (Figs. 4(d–f)). This indicates that the dielectric response of the low-frequency mode is not simply a local molecular rotation, but includes a collective polarization displacement whose amplitude increases with cell thickness. On the other hand, the relaxation frequency $f_{r,L}$ remains nearly constant and shows no significant thickness dependence. Thus, the restoring force that determines the relaxation time of the low-frequency mode is not governed by the macroscopic cell thickness, but rather by degrees of freedom intrinsic to the bulk LC.

This distinction becomes even clearer when examining the quantity $\Delta\varepsilon_{s-\infty,L}f_{r,L}$. As shown in Fig. 4(h), $\Delta\varepsilon_{s-\infty,L}f_{r,L}$ varies substantially with both thickness and alignment condition, and no universal scaling relation is observed. In contrast, $\Delta\varepsilon_{s-\infty,H}f_{r,H}$ remains nearly constant. Moreover, the temperature dependence shown in Fig. 4(i) reveals that $\Delta\varepsilon_{s-\infty,H}f_{r,H}$ remains approximately constant throughout the entire $N_F$ phase, whereas $\Delta\varepsilon_{s-\infty,L}f_{r,L}$ changes significantly with temperature. Therefore, the scaling behavior of the high-frequency mode is not limited to a particular temperature and represents a general characteristic of the $N_F$ phase. The temperature dependence of $\Delta\varepsilon_{s-\infty}$ and $f_r$ used to calculate $\Delta\varepsilon_{s-\infty}f_r$ for each alignment condition are shown in Fig. S6.

The influence of the alignment layer also differs markedly between the two modes. A particularly notable contrast is observed between the cells without an alignment layer and those coated with LIA-03. While the dielectric strength of the high-frequency mode is reduced by approximately a factor of 1.5 in the LIA-03-coated cell, the dielectric strength of the low-frequency mode is enhanced by nearly a factor of 5. Thus, the interfacial condition exerts opposite effects on the two modes. This contrasting behavior indicates that the two modes couple to the interface through fundamentally different mechanisms.

These results demonstrate that the high-frequency and low-frequency modes are different collective modes governed by different restoring forces. The high-frequency mode exhibits a clear thickness-dependent scaling relation characterized by $\Delta\varepsilon_{s-\infty,H}\, f_{r,H} \approx \text{const.}$, whereas no such scaling is observed for the low-frequency mode. Instead, the low-frequency mode exhibits distinct responses to thickness and alignment condition. Therefore, the two relaxation modes originate from fundamentally different collective dynamics.

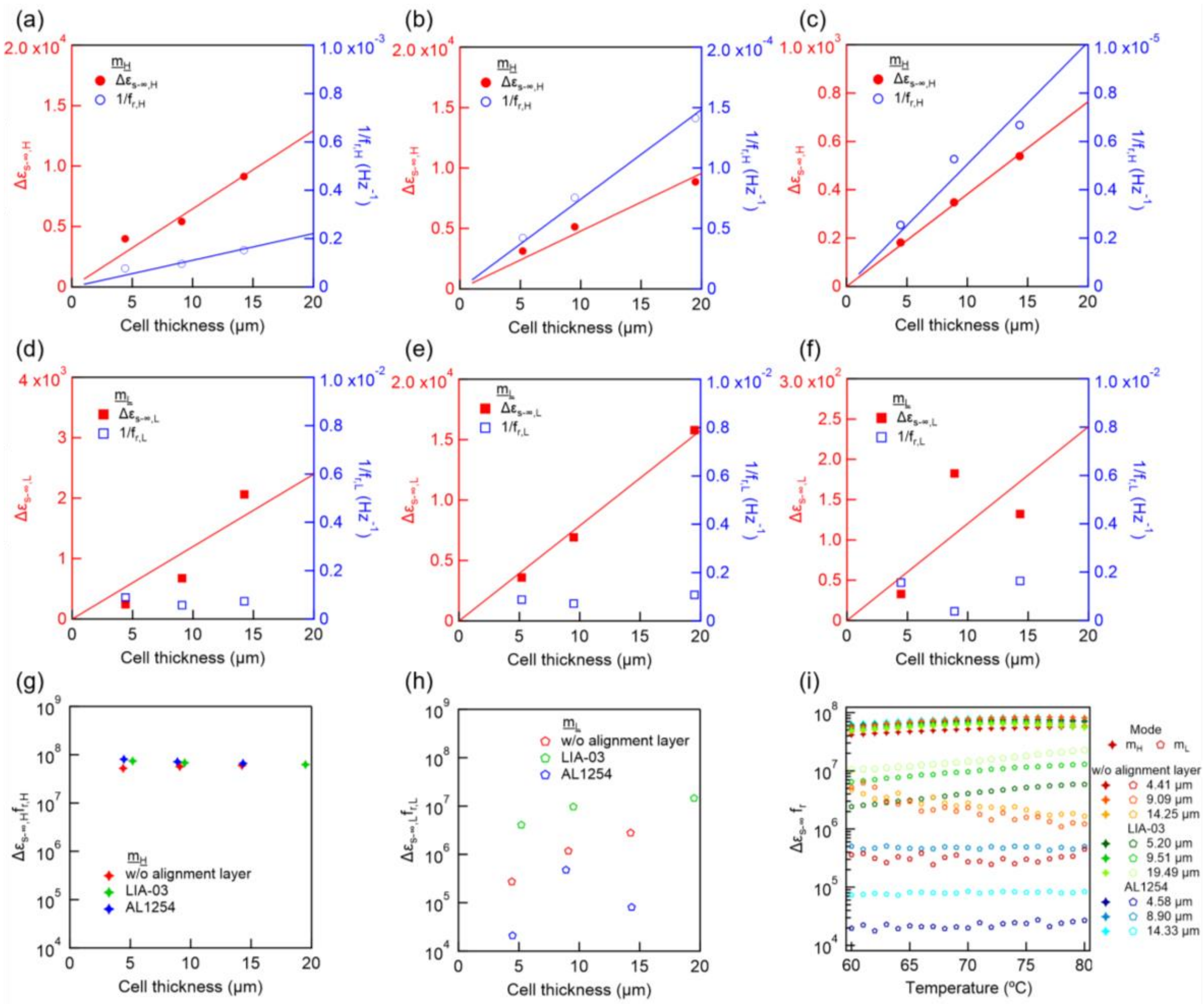

**Figure 4** Thickness and alignment dependence of the two relaxation modes. (a–c) Cell-thickness dependence of the dielectric strength ($\Delta\varepsilon_{s-\infty,H}$) (red symbols, left axis) and inverse relaxation frequency ($1/f_{r,H}$) (blue symbols, right axis) of the high-frequency mode for cells (a) without an alignment layer, (b) coated with LIA-03, and (c) coated with AL1254. Solid lines are linear fits passing through the origin. The dielectric strength increases approximately linearly with cell thickness,

whereas the relaxation frequency exhibits an inverse thickness dependence. (d–f) Corresponding thickness dependence of the dielectric strength ($\Delta\varepsilon_{s-\infty,L}$) and inverse relaxation frequency ($1/f_{r,L}$) of the low-frequency mode. While the dielectric strength increases with cell thickness, the relaxation frequency remains nearly independent of thickness. (g) Product of dielectric strength and relaxation frequency, $\Delta\varepsilon_{s-\infty,H}f_{r,H}$, for the high-frequency mode. (h) Corresponding values of $\Delta\varepsilon_{s-\infty,L}f_{r,L}$ for the low-frequency mode. (i) Temperature dependence of $\Delta\varepsilon_{s-\infty}f_r$ for the two relaxation modes measured in cells with different thicknesses and alignment conditions. Filled symbols correspond to the high-frequency mode ($m_H$) and open symbols to the low-frequency mode ($m_L$). The quantity $\Delta\varepsilon_{s-\infty,H}f_{r,H}$ remains nearly constant over the investigated thickness and temperature ranges, whereas $\Delta\varepsilon_{s-\infty,L}f_{r,L}$ exhibits a strong dependence on thickness, temperature, and alignment condition.

## 3. Discussion

### 3.1 Applicability and limitations of the PCG model

The results described above demonstrate that the giant dielectric response of the DIO mixture consists of at least two different relaxation modes. We now discuss the physical origins of these modes.

We first consider whether the observed dielectric response can be understood solely within a PCG-type picture. In previous studies, the giant dielectric permittivity of FNLCs has often been discussed in terms of the PCG model, where the dielectric response is attributed to polarization charge accumulation near interfaces. In this model, the apparent dielectric strength is governed mainly by the capacitance ratio between the LC layer and the insulating interfacial layer, approximately as

$$\Delta\varepsilon_{\mathrm{app}} \sim \frac{\varepsilon_i}{d_i} d_{\mathrm{LC}} \tag{2}$$

where $d_i$ and $\varepsilon_i$ are the thickness and permittivity of the interfacial layer, and $d_{\mathrm{LC}}$ is the LC thickness[25].

Our present results indicate that such a purely capacitive interpretation is insufficient, at least for the present material system. One indication is the alignment-layer dependence of the low-frequency real permittivity. As shown in Figs. 1(a,b), the low-frequency value of $\varepsilon'$ is larger in the LIA-03-coated cell than in the cell without an alignment layer. Within a simple interfacial-capacitance picture, introducing an additional insulating layer would increase the effective interfacial thickness and is therefore expected to reduce, rather than enhance, the apparent dielectric strength. Another indication is the electric-field dependence of the high-frequency mode. As shown in Fig. 3, $\Delta\varepsilon_{s-\infty,H}$ decreases markedly with increasing dc voltage. If the response were governed only by geometrical capacitance and interfacial permittivity, such a strong field-induced suppression of the dielectric strength would not be expected. These observations do not rule out interfacial contributions, but they show that the dielectric response cannot be described solely by a PCG-type capacitive mechanism. Instead, the collective motion of spontaneous polarization is more naturally described within a high-$\varepsilon$-type framework.

### 3.2 Low-frequency mode as a soft-mode-like short-axis rotational fluctuation

We next assign the two relaxation modes based on these experimental signatures. As discussed in Sec. 2.3, the low-frequency mode exhibits soft-mode-like behavior: the dielectric strength, $\Delta\varepsilon_{s-\infty,L}$, develops a pronounced peak near the SmZ$_A$–N$_F$ phase transition, and this response remains large even when the director is reoriented toward the electric-field direction. This indicates that the low-frequency mode is not a fluctuation around the director axis. Instead, it is more naturally assigned to a motion around an axis perpendicular to the director, namely molecular rotation around the short molecular axis, as schematically shown in Fig. 5(a).

This assignment is also consistent with the thickness and alignment dependences shown in Fig. 4. The relaxation frequency of the low-frequency mode $f_{r,L}$ is almost independent of the cell thickness and alignment layer, suggesting that the restoring force of this motion is governed mainly by a local change in the orientational order, rather than by a macroscopic elastic deformation over the entire cell. In this sense, the low-frequency mode reflects a soft-mode-like fluctuation of the ferroelectric order parameter. On the other hand, under zero dc field, the dielectric strength $\Delta\varepsilon_{s-\infty,L}$ depends on the cell thickness and alignment condition. Therefore, although the restoring force is essentially local, the dielectric amplitude observed at the cell level contains a collective polarization displacement coupled to the boundaries.

To describe this behavior, we consider a phenomenological free-energy model for the low-frequency mode. A more detailed derivation is given in Supplementary Note I. In the planar configuration, molecular rotation around the short molecular axis induces a displacement of the director and polarization toward the electric-field direction (Fig. 5(b)). Because this displacement is suppressed at the interfaces and becomes largest in the bulk, we approximate its spatial profile as

$$\phi_A(z) = A_L \sin\left(\frac{\pi z}{d}\right), \tag{3}$$

where $A_L$ is the maximum displacement amplitude at the cell center and $d$ is the cell thickness. This displacement is accompanied by a perturbation of the orientational distribution. We describe this effect using an orientational order parameter $\eta = \langle \cos^2\theta \rangle$ and assume that the displacement changes the local order parameter as

$$\eta(z) = \eta_0 - a\phi_A(z), \tag{4}$$

where $\eta_0$ is the equilibrium value and $a$ is a phenomenological coupling coefficient. The free-energy cost associated with this order-parameter distortion is then

$$F_{\mathrm{order}} = \int_0^d \frac{1}{2} K_\eta (\eta(z) - \eta_0)^2 dz = \int_0^d \frac{1}{2} K_\eta a^2 \phi_A(z)^2 dz. \tag{5}$$

Substituting the sinusoidal profile gives

$$F_{\mathrm{order}} = \frac{1}{4} K_0 A_L^2 d, \tag{6}$$

where $K_0 = K_\eta a^2$. Thus, $K_0$ represents the effective local restoring coefficient associated with the orientational-order distortion induced by short-axis molecular rotation.

We next include the interfacial displacement $\phi_s$, which represents the director or polarization displacement near the substrates. The interfacial displacement is coupled to the bulk amplitude $A_L$, and the corresponding interfacial free energy is written as

$$F_{\mathrm{int}} = w_\parallel \phi_s^2 + \frac{1}{2} d g (\phi_s - \lambda A_L)^2, \tag{7}$$

where $w_{\parallel}$ is the anchoring coefficient for the low-frequency displacement, $g$ is the coupling coefficient between the bulk and interfacial displacements, and $\lambda$ describes how efficiently the bulk displacement induces the interfacial displacement.

The electric field couples to both the spatially distributed bulk displacement and the interfacial displacement. The electric-field contribution is

$$F_E = -\int_0^d P_{\parallel} E(\phi_A(z) + \phi_s)dz = -P_{\parallel} Ed\left(\frac{2}{\pi}A_L + \phi_s\right), \tag{8}$$

where $P_{\parallel}$ is the polarization component associated with the short-axis rotational response.

The total free energy for the low-frequency mode is therefore

$$F_L(A_L, \phi_s) = \frac{1}{4}K_0 A_L^2 d + w_{\parallel}\phi_s^2 + \frac{1}{2}dg(\phi_s - \lambda A_L)^2 - P_{\parallel} Ed\left(\frac{2}{\pi}A_L + \phi_s\right). \tag{9}$$

To obtain the effective restoring force acting on the bulk amplitude $A_L$, we first eliminate the interfacial displacement by minimizing the free energy with respect to $\phi_s$:

$$\frac{\partial F_L}{\partial \phi_s} = 0. \tag{10}$$

This gives

$$\phi_s = \frac{dg\lambda}{2w_{\parallel} + dg}A_L + \frac{dP_{\parallel}}{2w_{\parallel} + dg}E. \tag{11}$$

Substituting this expression into the free energy, we obtain the effective free energy described only by $A_L$:

$$F_L(A_L) = \frac{1}{2}K_{\mathrm{eff},L}A_L^2 d - P_{\parallel} Ed\left(\frac{2}{\pi} + \frac{dg\lambda}{2w_{\parallel} + dg}\right)A_L - \frac{d^2 P_{\parallel}^2}{2(2w_{\parallel} + dg)}E^2, \tag{12}$$

where

$$K_{\mathrm{eff},L} = \frac{1}{2}K_0 + \frac{2gw_{\parallel}\lambda^2}{2w_{\parallel} + dg}. \tag{13}$$

The first term in $K_{\mathrm{eff},L}$ is the intrinsic restoring force associated with the local orientational-order distortion, whereas the second term arises from coupling to the interfacial anchoring. The equilibrium amplitude is obtained from

$$\frac{\partial F_L}{\partial A_L} = 0, \tag{14}$$

which gives

$$A_L = \frac{P_{\parallel} E\left(\frac{2}{\pi} + \frac{dg\lambda}{2w_{\parallel} + dg}\right)}{K_{\mathrm{eff},L}}. \tag{15}$$

The field-induced polarization response is evaluated from the spatial average of the induced polarization displacement:

$$\langle \delta P_L \rangle = \frac{1}{d}\int_0^d P_\parallel \left(\phi_A(z) + \phi_s\right) dz = P_\parallel \left(\frac{2}{\pi} A_L + \phi_s\right). \tag{16}$$

Substituting the expressions for $A_L$ and $\phi_s$, we obtain

$$\langle \delta P_L \rangle = P_\parallel^2 E \left[ \frac{\left(\frac{2}{\pi} + \frac{dg\lambda}{2w_\parallel + dg}\right)^2}{K_{\mathrm{eff},L}} + \frac{d}{2w_\parallel + dg} \right]. \tag{17}$$

Therefore, the susceptibility of the low-frequency mode is

$$\chi_L = \frac{\partial \langle \delta P_L \rangle}{\partial E} = P_\parallel^2 \left[ \frac{\left(\frac{2}{\pi} + \frac{dg\lambda}{2w_\parallel + dg}\right)^2}{K_{\mathrm{eff},L}} + \frac{d}{2w_\parallel + dg} \right], \tag{18}$$

and the dielectric strength is

$$\Delta\varepsilon_{s-\infty,L} \sim \frac{\chi_L}{\varepsilon_0}. \tag{19}$$

The relaxation dynamics are described by the overdamped equation of motion for $A_L$,

$$\gamma_L \frac{dA_L}{dt} = -K_{\mathrm{eff},L} A_L + P_\parallel E(t) \left(\frac{2}{\pi} + \frac{dg\lambda}{2w_\parallel + dg}\right), \tag{20}$$

where $\gamma_L$ is the effective rotational viscosity. Thus, the relaxation time is

$$\tau_L = \frac{\gamma_L}{K_{\mathrm{eff},L}}, \tag{21}$$

and the relaxation frequency is

$$f_{r,L} = \frac{K_{\mathrm{eff},L}}{2\pi\gamma_L}. \tag{22}$$

In the limit where the intrinsic orientational-order restoring force dominates over the interfacial contribution,

$$\frac{1}{2} K_0 \gg \frac{2gw_\parallel \lambda^2}{2w_\parallel + dg}, \tag{23}$$

the effective restoring coefficient becomes

$$K_{\mathrm{eff},L} \simeq \frac{1}{2} K_0 = \frac{1}{2} K_\eta a^2, \tag{24}$$

and therefore

$$f_{r,L} \simeq \frac{K_\eta a^2}{4\pi\gamma_L}. \tag{25}$$

In addition, when the interfacial contribution dominates the dielectric amplitude and the interface is sufficiently fixed, $2w_\parallel \gg dg$, the dielectric strength approximately becomes

$$\Delta\varepsilon_{s-\infty,L} \sim \frac{P_\parallel^2}{2\varepsilon_0 w_\parallel} d. \tag{26}$$

Thus, the low-frequency mode has two important characteristics: its relaxation frequency is governed mainly by the local orientational-order stiffness, whereas its dielectric strength can increase with cell thickness and depend on the interfacial condition. This explains why the low-frequency mode behaves as a soft-mode-like local fluctuation in its dynamics, while still appearing as a collective polarization response in the dielectric amplitude.

### 3.3 High-frequency mode as a Goldstone-like phase displacement of transverse polarization

In contrast, the high-frequency mode exhibits a relaxation frequency more than one order of magnitude larger than that of the low-frequency mode and is strongly suppressed by an applied dc electric field. Therefore, this mode is assigned to a collective motion around the director, as schematically illustrated in Fig. 5(c). It is well known that, in FNLC materials, the molecular dipole moment is generally tilted with respect to the molecular long axis.[38–40] Individual molecules retain some rotational freedom about their long axes. However, owing to dipole–dipole correlations, the azimuthal distribution of the tilted dipoles is not expected to be completely isotropic. As a result, an effective transverse polarization component can remain at the macroscopic level. The high-frequency mode is then interpreted as a collective phase displacement of this transverse polarization component around the director. Because this phase displacement corresponds to motion along an approximately degenerate azimuthal degree of freedom associated with the spontaneous breaking of global continuous rotational symmetry, the high-frequency mode can be regarded as a Goldstone-like collective mode.

We again formulate this mode using a phenomenological free energy. The detailed derivation is given in Supplementary Note II. The effective transverse polarization component is written as

$$P_{\perp} = P_s \sin\varphi, \tag{27}$$

where $P_s$ is the magnitude of the local polar order and $\varphi$ is the tilt angle of the molecular dipole from the director. The collective phase displacement of this transverse polarization is denoted by $\phi_A(z)$ as shown in Fig. 5(d). Since the displacement is constrained at the interfaces and becomes largest in the bulk, we approximate its spatial profile as

$$\phi_A(z) = A_H \sin\left(\frac{\pi z}{d}\right), \tag{28}$$

where $A_H$ is the maximum amplitude at the cell center.

Although this motion does not require an appreciable deformation of the director field, the collective transverse-polarization phase can have a weak bulk restoring energy. We write this contribution as

$$F_{\mathrm{pol}} = \int_0^d \frac{1}{2} K_P \phi_A(z)^2 dz, \tag{29}$$

where $K_P$ is the bulk stiffness associated with the transverse-polarization phase. Substituting the sinusoidal profile gives

$$F_{\mathrm{pol}} = \frac{1}{2} K_P A_H^2 \int_0^d \sin^2\left(\frac{\pi z}{d}\right) dz = \frac{1}{4} K_P A_H^2 d. \quad (30)$$

We also introduce an interfacial phase displacement $\phi_s$, which represents the transverse-polarization phase displacement near the substrate. The interfacial free energy is

$$F_{\mathrm{int}} = w_\perp \phi_s^2 + \frac{1}{2} dg(\phi_s - \lambda A_H)^2, \quad (31)$$

where $w_\perp$ is the anchoring coefficient for the transverse-polarization phase.

The electric field couples to the transverse polarization through the phase displacement:

$$F_E = -\int_0^d P_\perp E(\phi_A(z) + \phi_s) dz. \quad (32)$$

Substituting Eq. (28),

$$F_E = -P_\perp E d\left(\frac{2}{\pi} A_H + \phi_s\right). \quad (33)$$

The total free energy for the high-frequency mode is therefore

$$F_H(A_H, \phi_s) = \frac{1}{4} K_P A_H^2 d + w_\perp \phi_s^2 + \frac{1}{2} dg(\phi_s - \lambda A_H)^2 - P_\perp E d\left(\frac{2}{\pi} A_H + \phi_s\right). \quad (34)$$

The interfacial displacement is eliminated by minimizing the free energy with respect to $\phi_s$:

$$\frac{\partial F_H}{\partial \phi_s} = 0. \quad (35)$$

This gives

$$\phi_s = \frac{dg\lambda}{2w_\perp + dg} A_H + \frac{dP_\perp}{2w_\perp + dg} E. \quad (36)$$

Substituting this relation into Eq. (34) gives the effective free energy for $A_H$:

$$F_H(A_H) = \frac{1}{2} K_{\mathrm{eff},H} A_H^2 d - P_\perp E d A_H \left(\frac{2}{\pi} + \frac{dg\lambda}{2w_\perp + dg}\right) - \frac{d^2 P_\perp^2}{2(2w_\perp + dg)} E^2, \quad (37)$$

where

$$K_{\mathrm{eff},H} = \frac{1}{2} K_P + \frac{2gw_\perp \lambda^2}{2w_\perp + dg}. \quad (38)$$

Here, the first term is the weak bulk stiffness of the transverse-polarization phase, whereas the second term is the boundary-induced restoring contribution. The equilibrium amplitude is obtained from

$$\frac{\partial F_H}{\partial A_H} = 0, \quad (39)$$

which gives

$$A_H = \frac{P_\perp E}{K_{\mathrm{eff},H}} \left(\frac{2}{\pi} + \frac{dg\lambda}{2w_\perp + dg}\right). \quad (40)$$

The average polarization response is given by

$$\langle \delta P_H \rangle = \frac{1}{d}\int_0^d P_\perp (\phi_A(z) + \phi_s) dz = P_\perp \left(\frac{2}{\pi} A_H + \phi_s\right). \tag{41}$$

Substituting $A_H$ and $\phi_s$, we obtain

$$\langle \delta P_H \rangle = P_\perp^2 E \left[ \frac{\left(\frac{2}{\pi} + \frac{dg\lambda}{2w_\perp + dg}\right)^2}{K_{\mathrm{eff},H}} + \frac{d}{2w_\perp + dg} \right]. \tag{42}$$

Thus, the susceptibility and dielectric strength of the high-frequency mode are

$$\chi_H = \frac{\partial \langle \delta P_H \rangle}{\partial E} = P_\perp^2 \left[ \frac{\left(\frac{2}{\pi} + \frac{dg\lambda}{2w_\perp + dg}\right)^2}{K_{\mathrm{eff},H}} + \frac{d}{2w_\perp + dg} \right], \tag{43}$$

and

$$\Delta\varepsilon_{s-\infty,H} \sim \frac{\chi_H}{\varepsilon_0}. \tag{44}$$

The dynamics of the high-frequency mode are described by the overdamped equation

$$\gamma_H \frac{dA_H}{dt} = -K_{\mathrm{eff},H} A_H + P_\perp E(t) \left(\frac{2}{\pi} + \frac{dg\lambda}{2w_\perp + dg}\right), \tag{45}$$

where $\gamma_H$ is the effective viscosity for the collective transverse-polarization phase displacement. Therefore, the relaxation frequency is

$$f_{r,H} = \frac{K_{\mathrm{eff},H}}{2\pi\gamma_H}. \tag{46}$$

The experimentally observed thickness scaling is obtained in the limit where the bulk stiffness of the transverse-polarization phase is much smaller than the boundary-induced restoring contribution,

$$\frac{1}{2} K_P \ll \frac{2gw_\perp\lambda^2}{2w_\perp + dg}, \tag{47}$$

for sufficiently thick cells,

$$dg \gg 2w_\perp, \tag{48}$$

and where the amplitude-mediated contribution dominates the dielectric response,

$$\frac{\left(\frac{2}{\pi} + \frac{dg\lambda}{2w_\perp + dg}\right)^2}{K_{\mathrm{eff},H}} \gg \frac{d}{2w_\perp + dg}. \tag{49}$$

Under these conditions, we obtain

$$f_{r,H} \simeq \frac{w_\perp\lambda^2}{\pi\gamma_H}\frac{1}{d}, \tag{50}$$

and

$$\Delta\varepsilon_{s-\infty,H} \sim \frac{P_\perp^2 d}{2\varepsilon_0 w_\perp \lambda^2}\left(\frac{2}{\pi} + \lambda\right)^2. \tag{51}$$

Furthermore, their product is

$$\Delta\varepsilon_{s-\infty,H} f_{r,H} \simeq \frac{P_\perp^2}{2\pi\varepsilon_0\gamma_H}\left(\frac{2}{\pi}+\lambda\right)^2, \tag{52}$$

which is independent of both cell thickness and anchoring strength. This explains the nearly universal scaling of $\Delta\varepsilon_{s-\infty,H} f_{r,H}$ observed in Fig. 4.

Thus, the high-frequency mode is assigned to a collective phase displacement of the transverse polarization component around the director. This motion does not require an appreciable deformation of the director field, and therefore the Frank elastic contribution is negligible. The bulk transverse-polarization phase is nearly degenerate, while the finite restoring force mainly arises from boundary-induced symmetry breaking. In this sense, the high-frequency mode can be regarded as a Goldstone-like collective mode.

Taken together, these results show that the giant dielectric response of FNLCs is not a single relaxation process but a superposition of two collective polarization modes with different symmetries. The low-frequency mode is a soft-mode-like fluctuation associated with molecular rotation around the short molecular axis and local changes in the orientational order. The high-frequency mode is a Goldstone-like phase displacement of the transverse polarization component around the director. This assignment provides a unified interpretation of the temperature, electric-field, thickness, and alignment-layer dependences of the dielectric response.

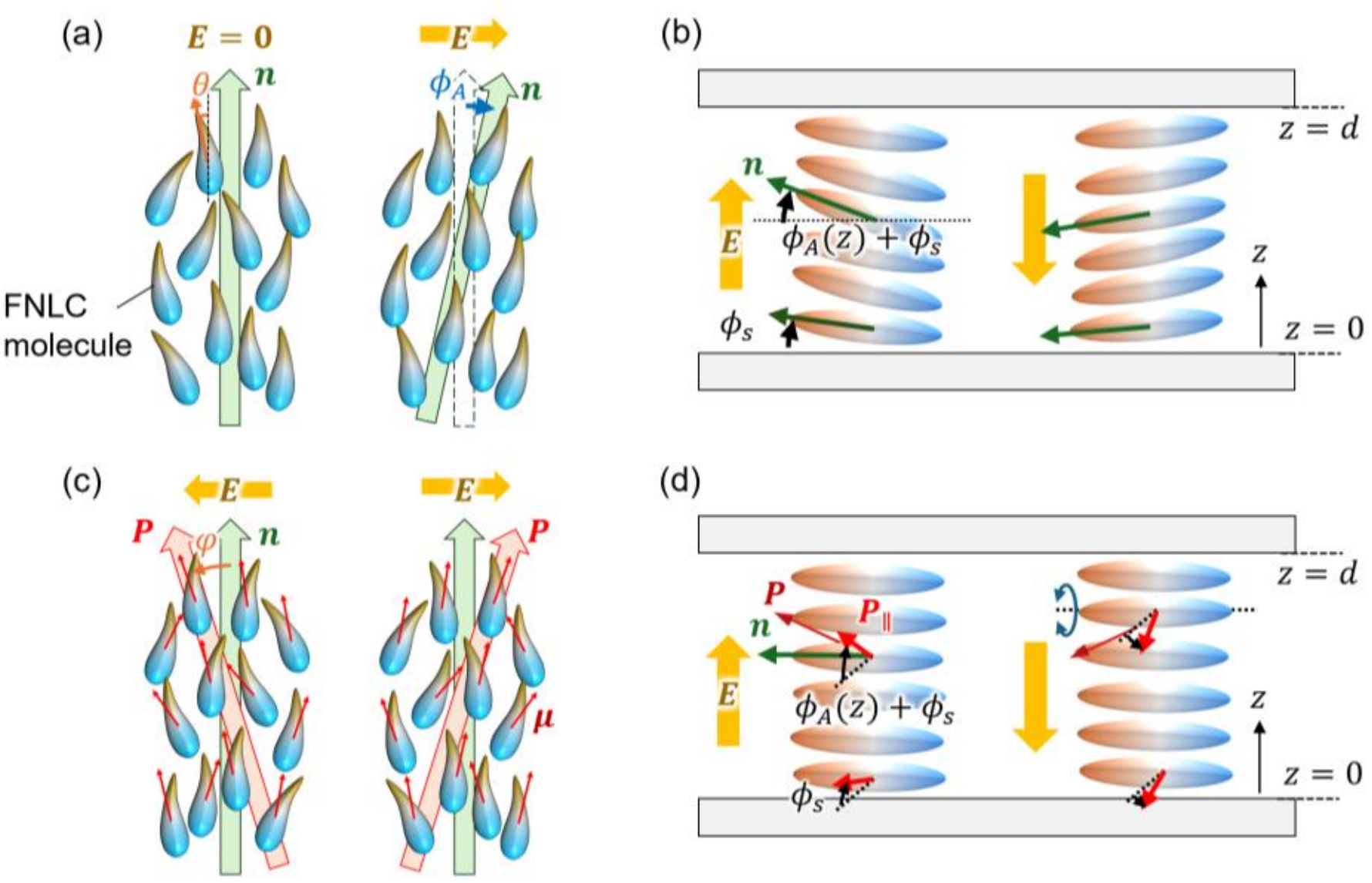

**Figure 5** Proposed collective modes underlying the giant dielectric response in the $N_F$ phase. (a,b) Low-frequency soft-mode-like relaxation. (a) Molecular rotation around the short molecular axis associated with local orientational-order fluctuations. (b) In a planar cell, this motion appears as a collective displacement of the director and polarization toward the electric-field direction. The displacement is described by a spatially distributed bulk component $\phi_A(z)$ and an interfacial

component $\phi_s$. (c,d) High-frequency Goldstone-like relaxation. (c) Collective molecular rotation around the director modulates the transverse polarization. Because the molecular dipole moment is tilted from the long axis, dipole–dipole correlations among the tilted dipoles can produce an effective transverse polarization component at the macroscopic level. (d) In a planar cell, this mode appears as a collective phase displacement of the transverse polarization around the director. The displacement is described by the bulk component $\phi_A(z)$ and the interfacial component $\phi_s$, and the effective transverse polarization component is denoted by $P_\perp$.

## 4. Conclusion

In this study, we revealed that the giant dielectric response of the DIO mixture arises from the superposition of two distinct collective modes rather than from a single relaxation process. This was demonstrated by systematically analyzing the temperature, dc electric-field, cell-thickness, and alignment-condition dependences of the dielectric spectra using Cole–Cole fitting and Bayesian model selection.

The low-frequency mode exhibits a pronounced anomaly near the $SmZ_A$–$N_F$ phase transition and is assigned to a soft-mode-like fluctuation associated with short-axis molecular rotation and ferroelectric order-parameter fluctuations. In contrast, the high-frequency mode is strongly suppressed by electric fields and follows an approximately universal relation, $\Delta\varepsilon_{s-\infty,H} f_{r,H} \approx \text{const.}$, supporting its interpretation as a Goldstone-like phase displacement of an effective transverse polarization component rotating around the director.

These findings provide a framework for decomposing the giant dielectric response of FNLCs into physically meaningful collective modes and deepen our understanding of polarization dynamics in polar liquid matter.

## 5. Methods

### 5.1 Materials

An FNLC mixture consisting of three fluoro-4-(3,4,5-trifluorophenyl)phenyl 2,6-difluoro-4-(trans-5-n-propyl-1,3-dioxane-2-yl)benzoate (DIO) derivatives with different alkyl-chain lengths was used in this study (Fig. S7). The mixture contained C1-DIO-35F-3F-345F, C2-DIO-35F-3F-345F, and C4-DIO-35F-3F-345F. The exact composition ratio is proprietary and therefore is not disclosed. The mixture exhibits a $SmZ_A$–$N_F$ phase transition near 84 °C.

### 5.2 Cell preparation

Sandwich cells were fabricated using two indium tin oxide (ITO)-coated glass substrates (Fig. S8). The electrode overlap area was 2 mm × 2 mm, corresponding to an active measurement area of 4 mm$^2$.

Three surface conditions were investigated: without an alignment layer, an azobenzene-based photoalignment layer (LIA-03, DIC), and a polyimide planar-alignment layer (AL1254, JSR). No rubbing treatment or photoalignment exposure was applied in this study in order to eliminate any predefined easy axis and to focus on the intrinsic dielectric response of the FNLC.

The cell gap was controlled using spacer particles and measured from the interference spectrum of the empty cells. Cell thicknesses ranging from approximately 4 to 20 μm were prepared for the thickness-dependence measurements.

### 5.3 Dielectric spectroscopy

Dielectric measurements were carried out using an impedance analyzer (4194A, YHP). A 20 mV $\mu m^{-1}$ AC probing electric field was applied while sweeping the frequency from 10 Hz to 1 MHz. For electric-field-dependent measurements, a dc bias voltage was superimposed on the AC probing signal.

The capacitance $C$ and conductance $G$ were measured and converted into the complex dielectric permittivity according to

$$\varepsilon'(\omega) - i\varepsilon''(\omega) = \frac{d}{\varepsilon_0 S}\left(C - \frac{iG}{\omega}\right), \tag{53}$$

where $S$ is the electrode area, $d$ is the cell thickness, and $\varepsilon_0$ is the vacuum permittivity.

Temperature was controlled using a Linkam hot stage (TS350, Linkam). Unless otherwise stated, measurements were performed during cooling at a rate of 0.2 $°\mathrm{C}\ \mathrm{min}^{-1}$. For dc-bias measurements, the voltage was repeatedly swept between 0 and 5 V eight times at 80 °C before data acquisition. The dielectric strength and relaxation frequencies were confirmed to converge after this conditioning procedure, indicating that transient effects associated with ionic adsorption at the interfaces had become negligible (Fig. S3).

### 5.4 Dielectric relaxation analysis

The dielectric spectra were analyzed using the Cole–Cole relaxation model. The complex dielectric permittivity was fitted using

$$\varepsilon^*(\omega) = \varepsilon_\infty + \sum_{i=1}^{N} \frac{\Delta\varepsilon_{s-\infty,i}}{1 + (i\omega\tau_i)^{1-\alpha_i}} - \frac{i\sigma_{\mathrm{DC}}}{\varepsilon_0\omega} \tag{54}$$

where $\varepsilon_\infty$ is the high-frequency permittivity, $\Delta\varepsilon_{s-\infty,i}$ is the dielectric strength, $\tau_i$ is the relaxation time, $\alpha_i$ is the Cole–Cole broadening parameter, and $\sigma_{\mathrm{DC}}$ is the dc conductivity.

Both one-relaxation (one-arc) and two-relaxation (two-arc) models were fitted to the measured spectra. The dc conductivity $\sigma_{\mathrm{DC}}$ was estimated from the low-frequency region and held constant during fitting. The high-frequency permittivity $\varepsilon_\infty$ was fixed at 10 to reduce the number of fitting parameters, while the dielectric strength $\Delta\varepsilon_{s-\infty,i}$, relaxation time $\tau_i$, and Cole–Cole broadening parameter $\alpha_i$ were optimized by nonlinear least-squares fitting. When $\varepsilon_\infty$ was treated as a free parameter, the fitting became unstable because of strong correlations among $\varepsilon_\infty$, the dielectric strengths, the Cole–Cole broadening parameters, and the dc-conductivity term. Therefore, $\varepsilon_\infty = 10$ was used as a representative high-frequency permittivity throughout the analysis.

To evaluate the statistical validity of the models, the Bayesian information criterion (BIC) was calculated as

$$\mathrm{BIC} = n \ln\left(\frac{\mathrm{RSS}}{n}\right) + k \ln n, \tag{55}$$

where $n$ is the number of residual data points, $k$ is the number of fitting parameters, and RSS is the residual sum of squares. The RSS was calculated from the combined residual vector of $\varepsilon'$ and $\varepsilon''$, using the same weighting scheme for the one-arc and two-arc models. The difference

$$\Delta\mathrm{BIC} = \mathrm{BIC}_{\mathrm{1arc}} - \mathrm{BIC}_{\mathrm{2arc}} \tag{56}$$

was used to compare the two models.

**Conflicts of interest**

There are no conflicts to declare.

**Acknowledgements**

The authors thank Prof. Go Watanabe for fruitful discussions. Partial components of the liquid crystal mixture were provided by JNC Petrochemical Corporation. This work was partly supported by MEXT KAKENHI (JP23H02038, JP23H00303, JP25K23593, and JP26K17651), Grant-in-Aid for JSPS Fellows (JP24KJ1622) and JST ACT-X (JPMJAX24DE).

## References

(1) Nishikawa, H.; Shiroshita, K.; Higuchi, H.; Okumura, Y.; Haseba, Y.; Yamamoto, S.; Sago, K.; Kikuchi, H. A Fluid Liquid-Crystal Material with Highly Polar Order. *Adv. Mater.* **2017**, *29* (43), 1702354. https://doi.org/10.1002/adma.201702354.

(2) Mandle, R. J.; Cowling, S. J.; Goodby, J. W. A Nematic to Nematic Transformation Exhibited by a Rod-like Liquid Crystal. *Phys. Chem. Chem. Phys.* **2017**, *19* (18), 11429–11435. https://doi.org/10.1039/C7CP00456G.

(3) Mandle, R. J.; Cowling, S. J.; Goodby, J. W. Rational Design of Rod-Like Liquid Crystals Exhibiting Two Nematic Phases. *Chem. – A Eur. J.* **2017**, *23* (58), 14554–14562. https://doi.org/10.1002/chem.201702742.

(4) Chen, X.; Korblova, E.; Dong, D.; Wei, X.; Shao, R.; Radzihovsky, L.; Glaser, M. A.; Maclennan, J. E.; Bedrov, D.; Walba, D. M.; Clark, N. A. First-Principles Experimental Demonstration of Ferroelectricity in a Thermotropic Nematic Liquid Crystal: Polar Domains and Striking Electro-Optics. *Proc. Natl. Acad. Sci.* **2020**, *117* (25), 14021–14031. https://doi.org/10.1073/pnas.2002290117.

(5) Sebastián, N.; Čopič, M.; Mertelj, A. Ferroelectric Nematic Liquid-Crystalline Phases. *Phys. Rev. E* **2022**, *106* (2), 1–27. https://doi.org/10.1103/PhysRevE.106.021001.

(6) Máthé, M. T.; Himel, M. S. H.; Adaka, A.; Gleeson, J. T.; Sprunt, S.; Salamon, P.; Jákli, A. Liquid Piezoelectric Materials: Linear Electromechanical Effect in Fluid Ferroelectric Nematic Liquid Crystals. *Adv. Funct. Mater.* **2024**, *34* (18), 2314158. https://doi.org/10.1002/adfm.202314158.

(7) Medle Rupnik, P.; Cmok, L.; Sebastián, N.; Mertelj, A. Viscous Mechano-Electric Response of Ferroelectric Nematic Liquid. *Adv. Funct. Mater.* **2024**, *34* (38), 2402554. https://doi.org/10.1002/adfm.202402554.

(8) Gill, M.; Máthé, M. T.; Salamon, P.; Gleeson, J. T.; Jákli, A. From Solid to Liquid Piezoelectric Materials. *Mater. Horiz.* **2025**, *12* (21), 8920–8942. https://doi.org/10.1039/D5MH00917K.

(9) Máthé, M. T.; Buka, Á.; Jákli, A.; Salamon, P. Ferroelectric Nematic Liquid Crystal Thermomotor. *Phys. Rev. E* **2022**, *105* (5), 1–6. https://doi.org/10.1103/PhysRevE.105.L052701.

(10) Adaka, A.; Guragain, P.; Perera, K.; Nepal, P.; Twieg, R. J.; Jákli, A. Low Field Electrocaloric Effect at Isotropic–Ferroelectric Nematic Phase Transition. *Soft Matter* **2025**, *21* (3), 458–462. https://doi.org/10.1039/D4SM00979G.

(11) Kavčič, A.; Sebastián, N.; Humar, M. Quasi-Phase-Matched Spontaneous Parametric down-Conversion in Chiral Ferroelectric Nematic Liquid Crystals. *Phys. Rev. Res.* **2026**, *8* (1), 013119. https://doi.org/10.1103/wgb3-clj5.

(12) Pan, J.-T.; Zhu, B.-H.; Ma, L.-L.; Chen, W.; Zhang, G.-Y.; Tang, J.; Liu, Y.; Wei, Y.; Zhang, C.; Zhu, Z.-H.; Zhu, W.-G.; Li, G.; Lu, Y.-Q.; Clark, N. A. Nonlinear Geometric Phase Coded Ferroelectric Nematic Fluids for Nonlinear Soft-Matter Photonics. *Nat. Commun.* **2024**, *15* (1),

8732. https://doi.org/10.1038/s41467-024-53040-8.

(13) Sebastián, N.; Lovšin, M.; Berteloot, B.; Osterman, N.; Petelin, A.; Mandle, R. J.; Aya, S.; Huang, M.; Drevenšek-Olenik, I.; Neyts, K.; Mertelj, A. Polarization Patterning in Ferroelectric Nematic Liquids via Flexoelectric Coupling. *Nat. Commun.* **2023**, *14* (1), 3029. https://doi.org/10.1038/s41467-023-38749-2.

(14) Folcia, C. L.; Ortega, J.; Vidal, R.; Sierra, T.; Etxebarria, J. The Ferroelectric Nematic Phase: An Optimum Liquid Crystal Candidate for Nonlinear Optics. *Liq. Cryst.* **2022**, *49* (6), 899–906. https://doi.org/10.1080/02678292.2022.2056927.

(15) Zhang, G.-Y.; Ma, L.-L.; Lin, E.; Wang, Z.-Y.; Pan, J.-T.; Yang, J.; Deng, M.; Wei, Y.; Ye, Y.; Wang, N.; Wang, Y.; Aya, S.; Lu, Y.-Q. Periodically-Modulated Unipolar and Bipolar Orders in Nematic Fluids towards Miniaturized Nonlinear Vectorial Optics. *Nat. Commun.* **2025**, *16* (1), 9419. https://doi.org/10.1038/s41467-025-64463-2.

(16) Blinc, R.; Zěkš, B. Dynamics of Helicoidal Ferroelectric Smectic-C̃ Liquid Crystals. *Phys. Rev. A* **1978**, *18* (2), 740–745. https://doi.org/10.1103/PhysRevA.18.740.

(17) Levstik, A.; Carlsson, T.; Filipic, C.; Levstik, I.; Zeks, B. Goldstone Mode and Soft Mode at the Smectic-A–Smectic-C* Phase Transition Studied by Dielectric Relaxation. *Phys. Rev. A* **1987**, *35* (8), 3527–3534. https://doi.org/10.1103/PhysRevA.35.3527.

(18) Carlsson, T.; Žekš, B.; Filipič, C.; Levstik, A. Theoretical Model of the Frequency and Temperature Dependence of the Complex Dielectric Constant of Ferroelectric Liquid Crystals near the Smectic-C̃. *Phys. Rev. A* **1990**, *42* (2), 877–889. https://doi.org/10.1103/PhysRevA.42.877.

(19) Yoshino, K.; Nakao, K.; Taniguchi, H.; Ozaki, M. Characteristic of Dielectric Behaviour of Ferroelectric Liquid Crystal at Smectic-A and Chiral Smectic-C Phase Transition. *J. Phys. Soc. Jpn.* **1987**, *56* (11), 4150–4156. https://doi.org/10.1143/JPSJ.56.4150.

(20) Ozaki, M.; Hatai, T.; Yoshino, K. Soft Mode Contribution around Sm A-Sm C* Phase Transition Temperature under DC Bias Field in Ferroelectric Liquid Crystal. *Jpn. J. Appl. Phys.* **1988**, *27* (11A), L1996. https://doi.org/10.1143/JJAP.27.L1996.

(21) Mandle, R. J.; Sebastián, N.; Martinez-Perdiguero, J.; Mertelj, A. On the Molecular Origins of the Ferroelectric Splay Nematic Phase. *Nat. Commun.* **2021**, *12* (1), 4962. https://doi.org/10.1038/s41467-021-25231-0.

(22) Ghimire, A.; Basnet, B.; Wang, H.; Guragain, P.; Baldwin, A.; Twieg, R.; Lavrentovich, O. D.; Gleeson, J.; Jakli, A.; Sprunt, S. Dynamics of the Antiferroelectric Smectic-$Z_A$ Phase in a Ferroelectric Nematic Liquid Crystal. *Soft Matter* **2025**, *21* (44), 8510–8522. https://doi.org/10.1039/D5SM00796H.

(23) Kats, E. I. Stability of the Uniform Ferroelectric Nematic Phase. *Phys. Rev. E* **2021**, *103* (1), 1–5. https://doi.org/10.1103/PhysRevE.103.012704.

(24) Sebastián, N.; Cmok, L.; Mandle, R. J.; De La Fuente, M. R.; Drevenšek Olenik, I.; Čopič, M.;

Mertelj, A. Ferroelectric-Ferroelastic Phase Transition in a Nematic Liquid Crystal. *Phys. Rev. Lett.* **2020**, *124* (3), 1–6. https://doi.org/10.1103/PhysRevLett.124.037801.

(25) Clark, N. A.; Chen, X.; MacLennan, J. E.; Glaser, M. A. Dielectric Spectroscopy of Ferroelectric Nematic Liquid Crystals: Measuring the Capacitance of Insulating Interfacial Layers. *Phys. Rev. Res.* **2024**, *6* (1), 013195. https://doi.org/10.1103/PhysRevResearch.6.013195.

(26) Erkoreka, A.; Martinez-Perdiguero, J.; Mandle, R. J.; Mertelj, A.; Sebastián, N. Dielectric Spectroscopy of a Ferroelectric Nematic Liquid Crystal and the Effect of the Sample Thickness. *J. Mol. Liq.* **2023**, *387* (March), 122566. https://doi.org/10.1016/j.molliq.2023.122566.

(27) Adaka, A.; Rajabi, M.; Haputhantrige, N.; Sprunt, S.; Lavrentovich, O. D.; Jákli, A. Dielectric Properties of a Ferroelectric Nematic Material: Quantitative Test of the Polarization-Capacitance Goldstone Mode. *Phys. Rev. Lett.* **2024**, *133* (3), 38101. https://doi.org/10.1103/PhysRevLett.133.038101.

(28) Erkoreka, A.; Martinez-Perdiguero, J. Constraining the Value of the Dielectric Constant of the Ferroelectric Nematic Phase. *Phys. Rev. E* **2024**, *110* (2), L022701. https://doi.org/10.1103/PhysRevE.110.L022701.

(29) Matko, V.; Gorecka, E.; Pociecha, D.; Matraszek, J.; Vaupotič, N. Interpretation of Dielectric Spectroscopy Measurements of Ferroelectric Nematic Liquid Crystals. *Phys. Rev. Res.* **2024**, *6* (4), L042017. https://doi.org/10.1103/PhysRevResearch.6.L042017.

(30) Vaupotič, N.; Krajnc, T.; Gorecka, E.; Pociecha, D.; Matko, V. Ferroelectric Nematics: Materials with High Permittivity or Low Resistivity? *Liq. Cryst.* **2025**, *2* (00), 1–13. https://doi.org/10.1080/02678292.2025.2484234.

(31) Yadav, N.; Panarin, Y. P.; Vij, J. K.; Jiang, W.; Mehl, G. H. Two Mechanisms for the Formation of the Ferronematic Phase Studied by Dielectric Spectroscopy. *J. Mol. Liq.* **2023**, *378*, 121570. https://doi.org/10.1016/j.molliq.2023.121570.

(32) Erkoreka, A.; Mertelj, A.; Huang, M.; Aya, S.; Sebastián, N.; Martinez-Perdiguero, J. Collective and Non-Collective Molecular Dynamics in a Ferroelectric Nematic Liquid Crystal Studied by Broadband Dielectric Spectroscopy. *J. Chem. Phys.* **2023**, *159* (18). 184502. https://doi.org/10.1063/5.0173813.

(33) Yoshino, K.; Nakao, K.; Taniguchi, H.; Ozaki, M. Influence of Electric Field and Cell Thickness on Dielectric Behaviour of Ferroelectric Liquid Crystal at Phase Transition. *Jpn. J. Appl. Phys.* **1987**, *26* (S2), 97. https://doi.org/10.7567/JJAPS.26S2.97.

(34) Vallerien, S. U.; Kremer, F.; Kapitza, H.; Zentel, R.; Frank, W. Field Dependent Soft and Goldstone Mode in a Ferroelectric Liquid Crystal as Studied by Dielectric Spectroscopy. *Phys. Lett. A* **1989**, *138* (4–5), 219–222. https://doi.org/10.1016/0375-9601(89)90032-7.

(35) Glogarová, M.; Pavel, J. The Effect of Biasing Electric Field on the Soft Mode in the Vicinity of the Ferroelectric Phase Transition in Liquid Crystals. *Liq. Cryst.* **1989**, *6* (3), 325–332.

https://doi.org/10.1080/02678298908029083.

(36) Nakase, M.; Kamifuji, H.; Nakajima, K.; Kikuchi, H.; Fukuda, K.; Ozaki, M. Field-Induced Phase Transition Behaviour of Ferroelectric Nematic Liquid Crystals under DC Electric Fields. *Soft Matter* **2026**, *22* (13), 2538–2544. https://doi.org/10.1039/D5SM01272D.

(37) Vaupotič, N.; Pociecha, D.; Rybak, P.; Matraszek, J.; Čepič, M.; Wolska, J. M.; Gorecka, E. Dielectric Response of a Ferroelectric Nematic Liquid Crystalline Phase in Thin Cells. *Liq. Cryst.* **2023**, *50* (4), 584–595. https://doi.org/10.1080/02678292.2023.2180099.

(38) Li, J.; Nishikawa, H.; Kougo, J.; Zhou, J.; Dai, S.; Tang, W.; Zhao, X.; Hisai, Y.; Huang, M.; Aya, S. Development of Ferroelectric Nematic Fluids with Giant-$\varepsilon$ Dielectricity and Nonlinear Optical Properties. *Sci. Adv.* **2021**, *7* (17), eabf5047. https://doi.org/10.1126/sciadv.abf5047.

(39) Sebastián, N.; Mandle, R. J.; Petelin, A.; Eremin, A.; Mertelj, A. Electrooptics of Mm-Scale Polar Domains in the Ferroelectric Nematic Phase. *Liq. Cryst.* **2021**, *48* (14), 2055–2071. https://doi.org/10.1080/02678292.2021.1955417.

(40) Matsukizono, H.; Iwamatsu, K.; Endo, S.; Okumura, Y.; Anan, S.; Kikuchi, H. Synthesis of Liquid Crystals Bearing 1,3-Dioxane Structures and Characterization of Their Ferroelectricity in the Nematic Phase. *J. Mater. Chem. C* **2023**, *11* (18), 6183–6190. https://doi.org/10.1039/D2TC05363B.

# Supplemental Information

## Uncovering Collective Modes Underlying the Giant Dielectric Response of Ferroelectric Nematic Liquid Crystals

Kazuma Nakajima[1*], Hirokazu Kamifuji[1], Masanori Ozaki[1], Hirotsugu Kikuchi[2], and Kenjiro Fukuda[1*]

[1]Division of Electrical, Electronic and Infocommunications Engineering, Graduate School of Engineering, The University of Osaka, 2-1 Yamadaoka, Suita, Osaka 565-0871, Japan

[2]Institute for Materials Chemistry and Engineering, Kyushu University, Kasuga, Fukuoka 816-8580, Japan

*nakajima.kazuma@eei.eng.osaka-u.ac.jp, fukuda@eei.eng.osaka-u.ac.jp

**Contents**

**Supplementary Figures**

**Supplementary Notes**

This PDF file includes Supplementary Figures S1–S8 and Supplementary Notes 1–2.

## Supplementary Figures

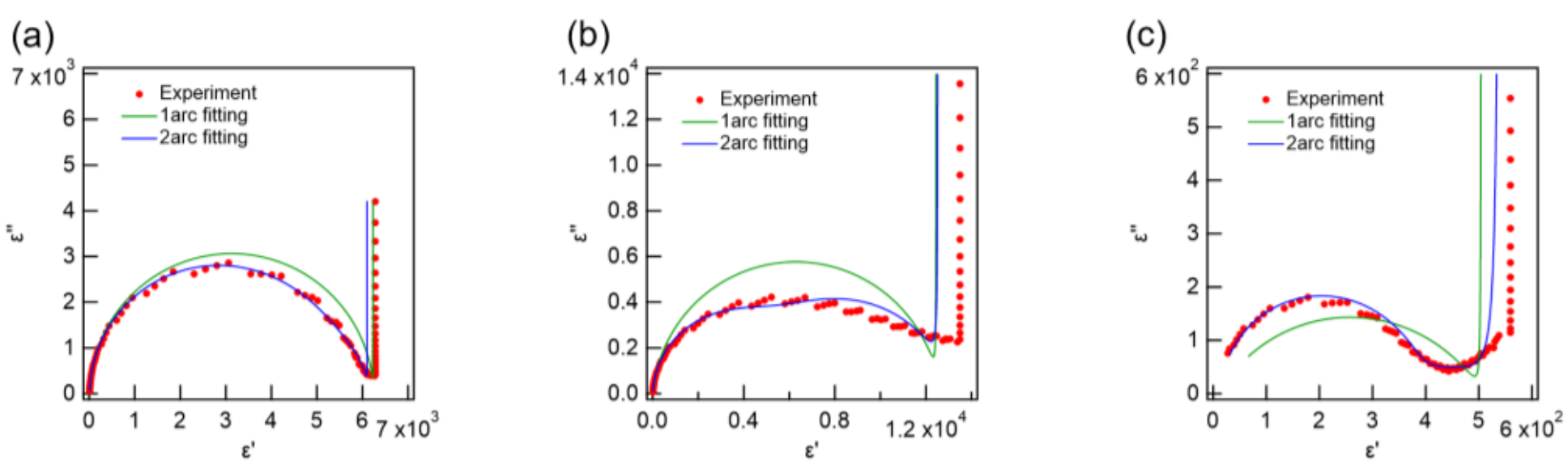

**Figure S1** Cole–Cole plots and fitting results for the ferroelectric nematic liquid crystal. (a–c) Cole–Cole plots measured at 75°C for (a) a 9.05-μm-thick cell without an alignment layer, (b) a 9.51-μm-LIA-03-coated cell, and (c) an 8.90-μm-AL1254 coated cell. Red symbols represent the experimental data. Green and blue curves correspond to the fitting results obtained using the one-relaxation (1 arc) and two-relaxation (2 arc) models, respectively. The 2 arc model reproduces the experimental Cole–Cole plots more accurately than the conventional 1 arc model, supporting the existence of two independent dielectric relaxation modes in the FNLC.

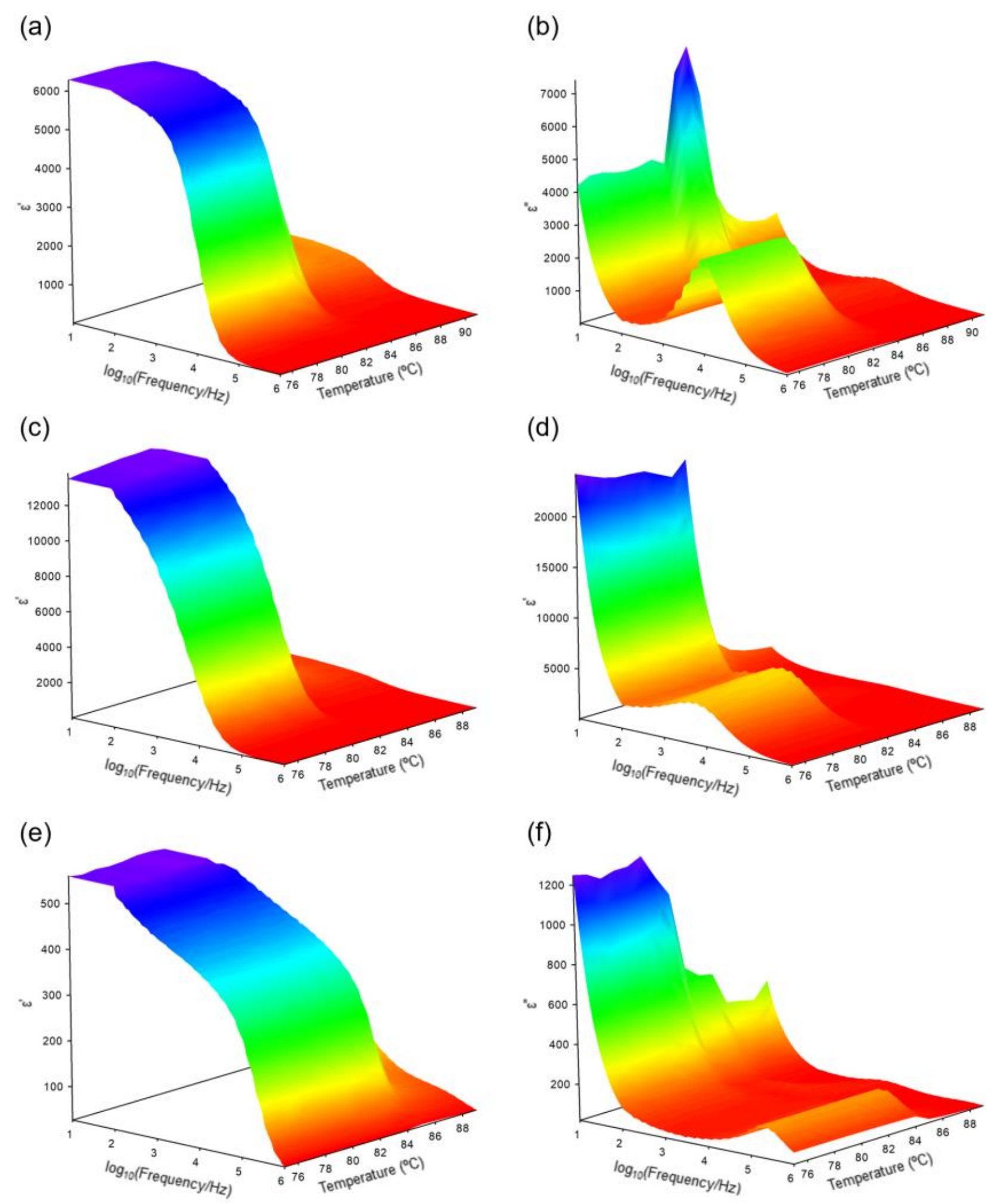

**Figure S2** Temperature evolution of the dielectric spectra used for the relaxation analysis. Three-dimensional plots of the real part and the imaginary part of the complex dielectric permittivity ($\varepsilon'$, $\varepsilon''$) as a function of frequency and temperature during cooling. (a,b) a 9.05-μm-thick cell without an alignment layer, (c,d) a 9.51-μm-LIA-03-coated cell, and (e,f) an 8.90-μm-AL1254 coated cell. Panels (a,c,e) show $\varepsilon'$, while panels (b,d,f) show $\varepsilon''$. The measurements were performed during cooling at a rate of $0.2\,°\mathrm{C\,min^{-1}}$. These data were used to extract the dielectric strengths and relaxation frequencies presented in Fig. 2.

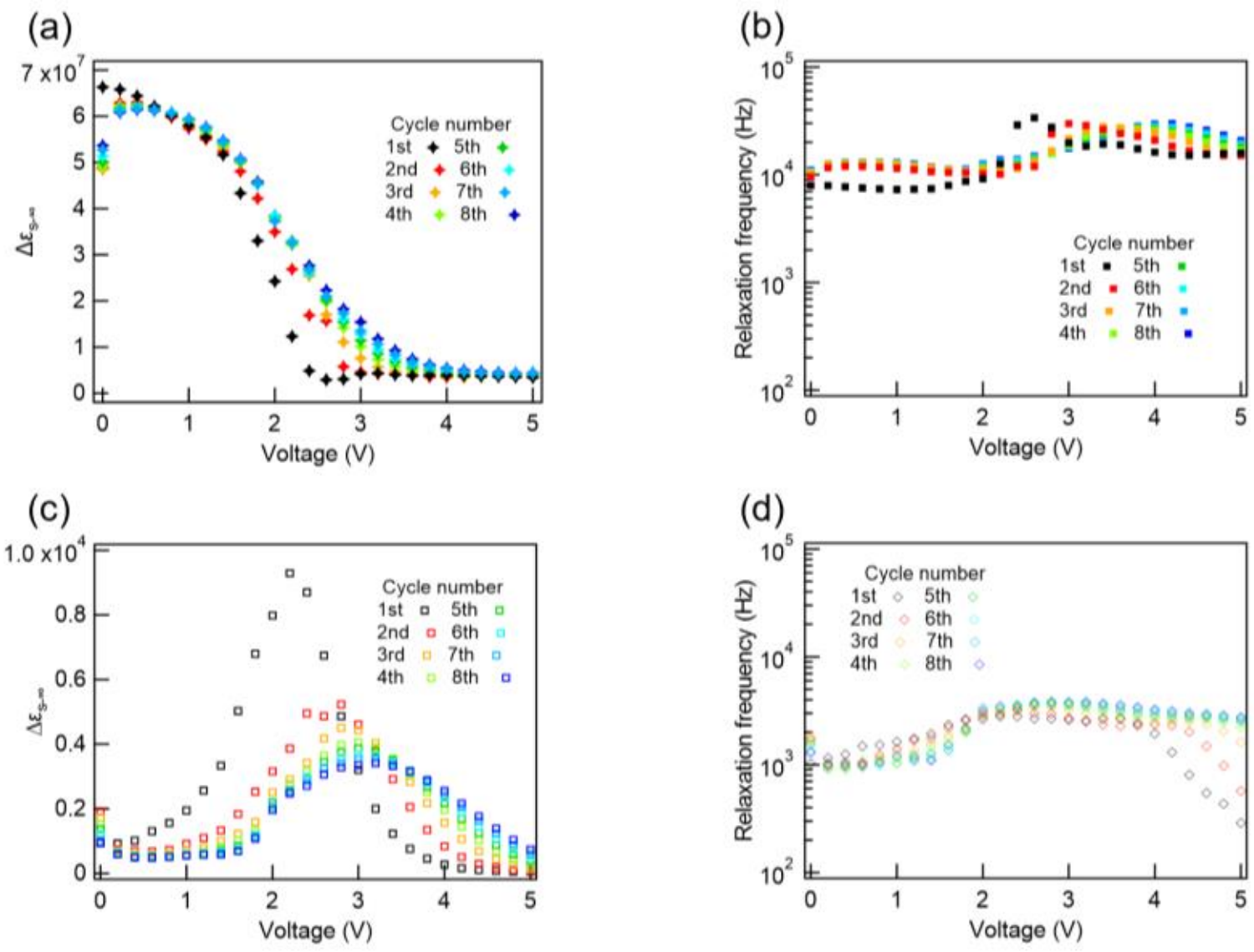

**Figure S3** Convergence of dielectric relaxation parameters during repeated dc-voltage cycling. Voltage dependence of the dielectric strength and relaxation frequency obtained from the 2 arc analysis at 80°C for a 14.25 μm cell without an alignment layer. (a) Dielectric strength $\Delta\varepsilon_{S-\infty,H}$ and (b) relaxation frequency $f_{r,H}$ of the high-frequency mode. (c) Dielectric strength $\Delta\varepsilon_{S-\infty,L}$ and (d) relaxation frequency $f_{r,L}$ of the low-frequency mode. The DC voltage was swept repeatedly from 0 to 5 V for 8 cycles. Significant changes are observed during the first few cycles, whereas the dielectric parameters converge after repeated voltage cycling. The measurements presented in Fig. 3(a–e) were performed after confirming the convergence of the dielectric response.

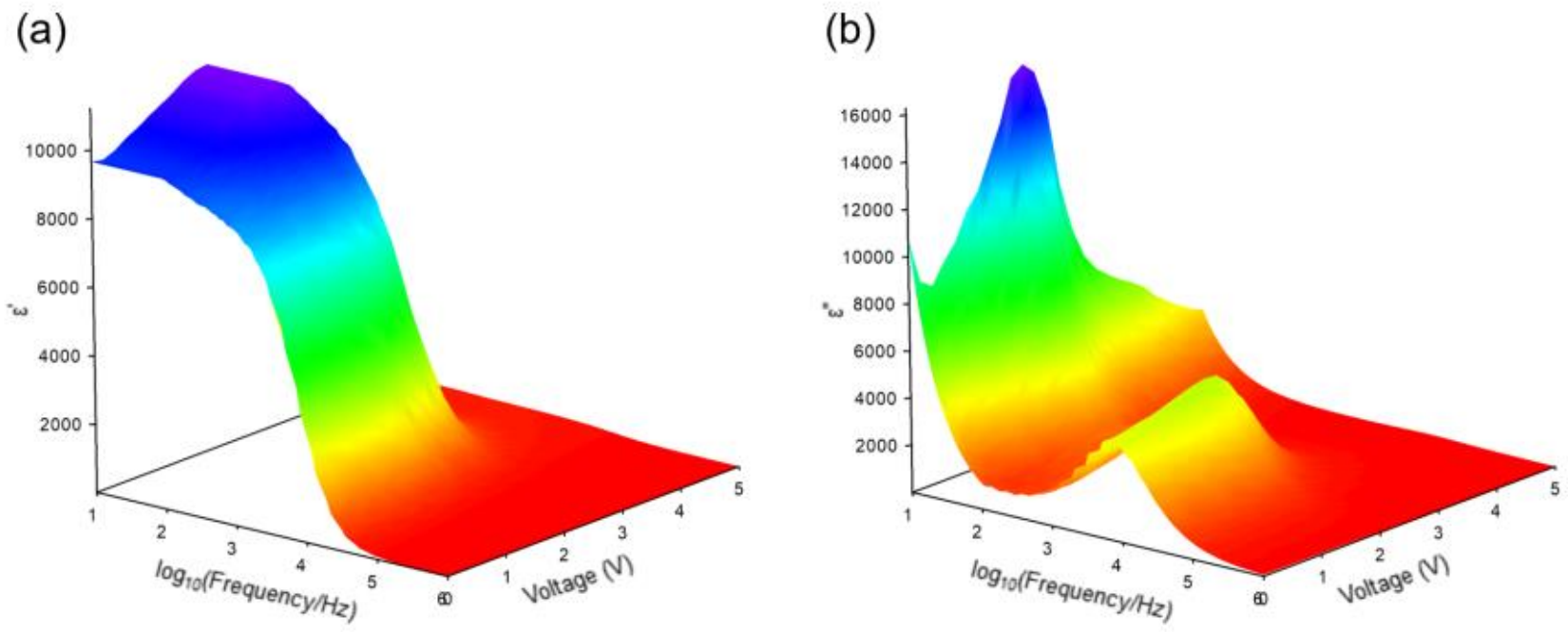

**Figure S4** Voltage dependence of the dielectric spectra in the ferroelectric nematic phase. Three-dimensional plots of (a) the real part and (b) the imaginary part of the complex dielectric permittivity as a function of frequency and applied DC voltage measured at 80°C in a 14.25 μm cell without an alignment layer. With increasing DC voltage, the dielectric-loss peak shifts toward lower frequencies and the contribution of the low-frequency relaxation mode becomes increasingly pronounced. These spectra were used to obtain the voltage dependences of the dielectric strengths and relaxation frequencies shown in Fig. 3(d,e).

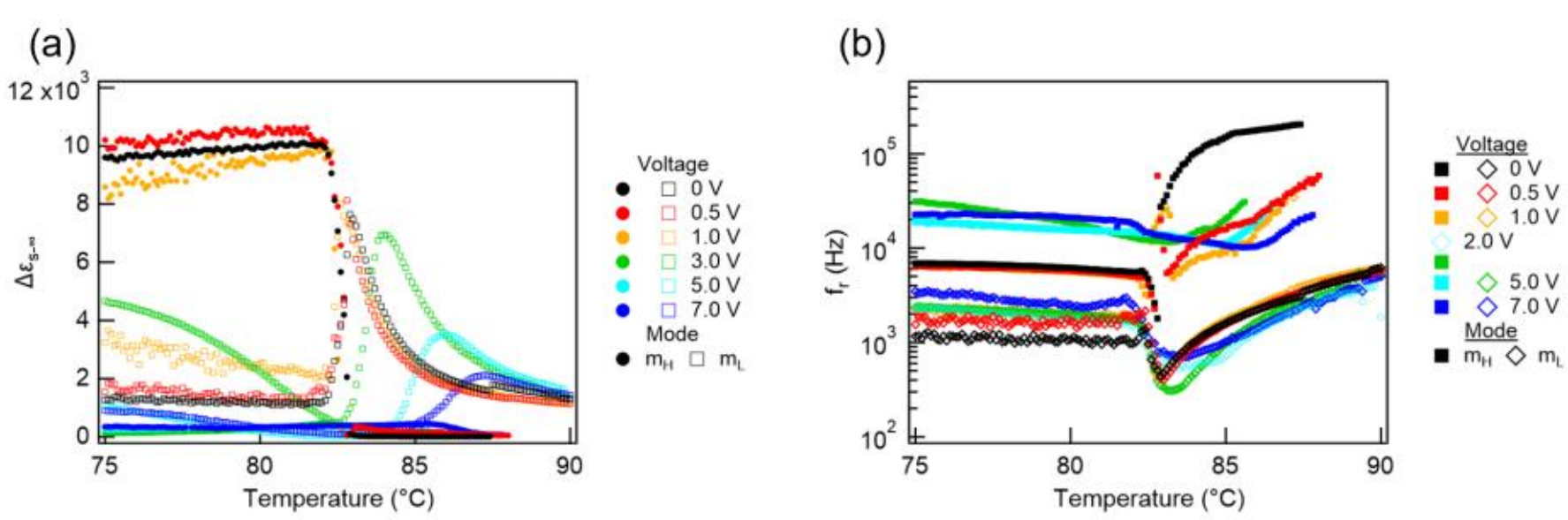

**Figure S5** Temperature dependence of the dielectric strengths and relaxation frequencies under applied dc voltages. (a) Temperature dependence of the dielectric strength $\Delta\varepsilon_{s-\infty}$ and (b) relaxation frequency $f_r$ obtained from the 2 arc analysis for different DC voltages in a 14.31 μm cell without an alignment layer. Filled symbols correspond to the high-frequency mode ($m_H$) and open symbols to the low-frequency mode ($m_L$). Increasing DC voltage suppresses the dielectric strength of the high-frequency mode, while the low-frequency mode remains pronounced and develops a peak that shifts toward higher temperature. Also, the minimum value of the relaxation frequency in the low-frequency mode was observed to shift toward higher temperatures when a voltage was applied.

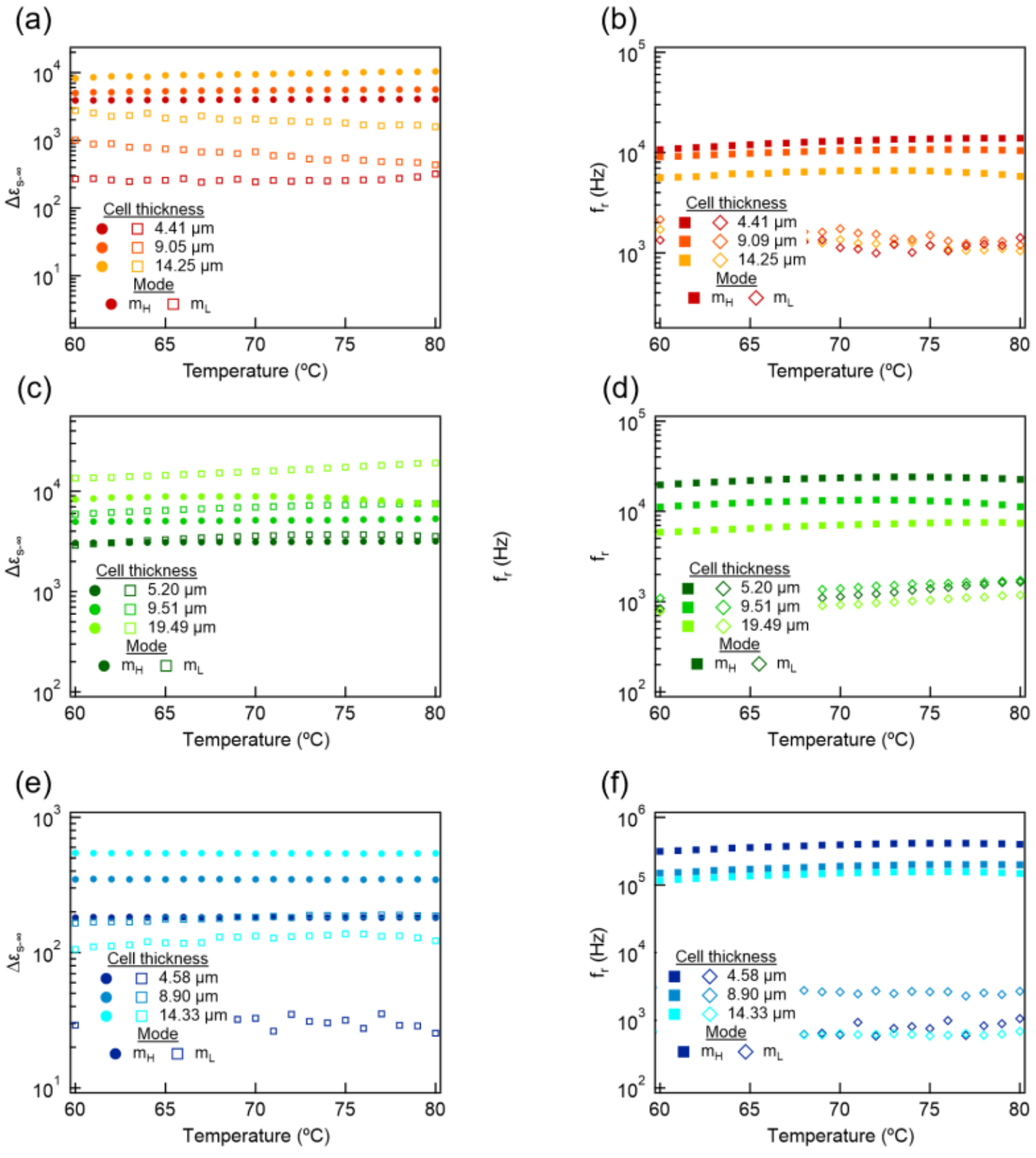

**Figure S6** Temperature dependences of the dielectric strengths and relaxation frequencies for different cell thicknesses. (a,b) Cells without an alignment layer (($d$ = 4.41, 9.05, and 14.25 μm), (c,d) LIA-03-coated cells ($d$ = 5.20, 9.51, and 19.49 μm), and (e,f) AL1254-coated cells ($d$ = 4.58, 8.90, and 14.33 μm). Panels (a,c,e) show the dielectric strength $\Delta\varepsilon_{s-\infty}$, and panels (b,d,f) show the relaxation frequency $f_r$. Filled symbols correspond to the high-frequency mode ($m_H$), and open symbols correspond to the low-frequency mode ($m_L$). These data were used for the thickness-scaling analysis presented in Fig. 4.

C1-DIO-35F-3F-345F

C2-DIO-35F-3F-345F

C4-DIO-35F-3F-345F

**Figure S7** Chemical structures of the FNLC components used in this study. Molecular structures of C1-DIO-35F-3F-345F, C2-DIO-35F-3F-345F, and C4-DIO-35F-3F-345F. The FNLC investigated in this work was prepared from a mixture of these compounds with different terminal alkyl-chain lengths.

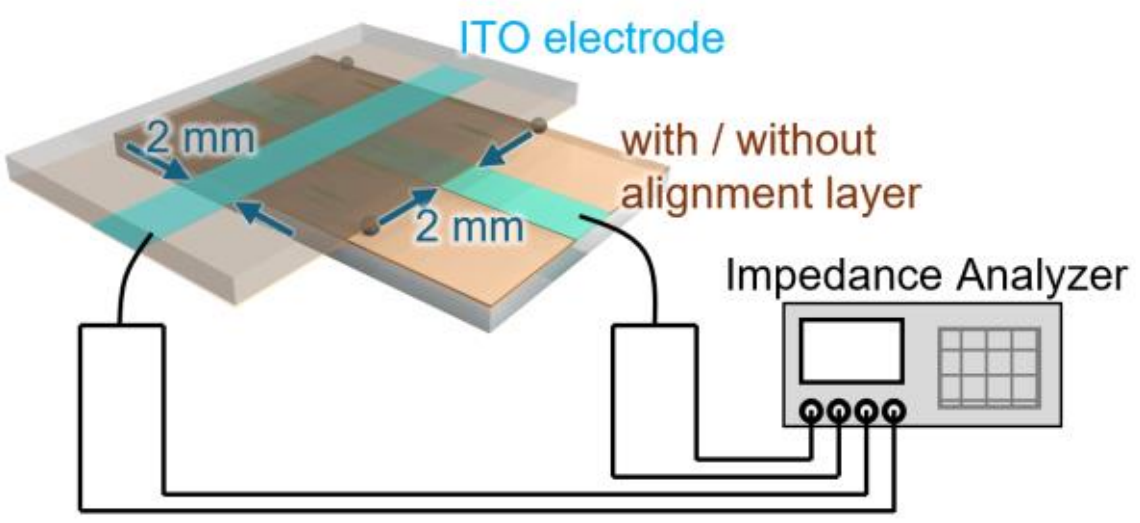

**Figure S8** Schematic illustration of the dielectric spectroscopy measurement setup. Two ITO-coated glass substrates were crossed orthogonally to form a measurement area of $2 \times 2\ \mathrm{mm}^2$. Cells were prepared with or without an alignment layer, and the complex dielectric permittivity was measured using an impedance analyzer. The cell thickness was controlled by spacer particles dispersed between the substrates.

# Supplementary Notes

**Supplementary Note I. Derivation of the low-frequency mode**

We describe the low-frequency mode as a soft-mode-like fluctuation associated with molecular rotation around the short molecular axis. In a planar cell, this motion is accompanied by a spatially distributed director displacement that is constrained by the interfaces. In contrast, when the director is reoriented toward the electric-field direction, the response is dominated by a more local change in the orientational order. We formulate these two limiting cases separately.

**I-1 Planar director configuration**

We first consider the case in which the polar director lies parallel to the substrates and the electric field is applied along the cell normal. In this configuration, molecular rotation around the short molecular axis induces a displacement of the director and of the orientational distribution toward the electric-field direction. Because the interfaces suppress this displacement, whereas the bulk region can deform more freely, the displacement is expected to be largest near the center of the cell. We approximate the spatial profile as

$$\phi_A(z) = A_L \sin\left(\frac{\pi z}{d}\right), \tag{S1}$$

where $A_L$ is the maximum displacement amplitude at $z = d/2$, and $d$ is the cell thickness.

The intrinsic restoring force of this mode is assumed to originate from changes in the orientational distribution around the director. We introduce the orientational order parameter

$$\eta = \langle \cos^2 \theta \rangle, \tag{S2}$$

where $\theta$ is the angle between the molecular long axis and the local director. The equilibrium value is denoted by $\eta_0$. A short-axis rotational displacement is accompanied by a small change in the orientational distribution, which we write phenomenologically as

$$\eta(z) - \eta_0 = a\phi_A(z), \tag{S3}$$

where $a$ is a coupling coefficient. The free-energy cost associated with this order-parameter distortion is

$$F_{\text{order}} = \int_0^d \frac{1}{2} K_\eta (\eta(z) - \eta_0)^2 dz, \tag{S4}$$

where $K_\eta$ is the stiffness of the orientational distribution. Substituting Eq. (S3) gives

$$F_{\text{order}} = \int_0^d \frac{1}{2} K_\eta a^2 \phi_A(z)^2 dz. \tag{S5}$$

Using Eq. (S1),

$$F_{\text{order}} = \frac{1}{2} K_\eta a^2 A_L{}^2 \int_0^d \sin^2\left(\frac{\pi z}{d}\right) dz. \tag{S6}$$

Since

$$\int_0^d \sin^2\left(\frac{\pi z}{d}\right) dz = \frac{d}{2}, \quad \text{(S7)}$$

we obtain

$$F_{\mathrm{order}} = \frac{1}{4} K_0 {A_L}^2 d, \quad \text{(S8)}$$

where

$$K_0 = K_\eta a^2. \quad \text{(S9)}$$

Thus, the coefficient $K_0$ represents the effective bulk restoring coefficient associated with the orientational-order distortion induced by the short-axis rotational mode.

Next, we include the interfacial displacement $\phi_s$, which represents the director displacement near the substrates. The interfacial displacement is coupled to the bulk amplitude $A_L$, and the corresponding free energy is written as

$$F_{\mathrm{int}} = w_\parallel \phi_s^2 + \int_0^d \frac{1}{2} g(\phi_s - \lambda A_L)^2 dz = w_\parallel \phi_s^2 + \frac{1}{2} dg(\phi_s - \lambda A_L)^2, \quad \text{(S10)}$$

where $w_\parallel$ is the anchoring coefficient for the director deformation associated with the low-frequency mode, $g$ is the coupling coefficient between the bulk and interfacial displacements, and $\lambda$ describes how efficiently the bulk distortion induces the interfacial displacement.

The electric field couples to both the spatially distributed bulk displacement and the interfacial displacement.

$$F_E = -\int_0^d P_\parallel E\big(\phi_s + \phi_A(z)\big) dz, \quad \text{(S11)}$$

where $P_\parallel$ is the polarization component associated with the short-axis rotational response. Using Eq. (S1),

$$F_E = -\left(P_\parallel E \phi_s d + P_\parallel E A_L \int_0^d \sin\left(\frac{\pi z}{d}\right) dz\right) = -P_\parallel E d\left(\frac{2}{\pi} A_L + \phi_s\right). \quad \text{(S12)}$$

Thus, the total free energy in the planar configuration is therefore

$$F_{\mathrm{pl}}(A_L, \phi_s) = \frac{1}{4} K_0 {A_L}^2 d + w_\parallel \phi_s^2 + \frac{1}{2} dg(\phi_s - \lambda A_L)^2 - P_\parallel E d\left(\frac{2}{\pi} A_L + \phi_s\right). \quad \text{(S13)}$$

The equilibrium interfacial displacement is obtained from

$$\frac{\partial F_{\mathrm{pl}}}{\partial \phi_s} = 0. \quad \text{(S14)}$$

This gives

$$2 w_\parallel \phi_s + dg(\phi_s - \lambda A_L) - P_\parallel E d = 0, \quad \text{(S15)}$$

and hence

$$\phi_s = \frac{dg\lambda}{2 w_\parallel + dg} A_L + \frac{d P_\parallel}{2 w_\parallel + dg} E. \quad \text{(S16)}$$

Substituting Eq. (S16) into Eq. (S13), the effective free energy for $A_L$ becomes

$$F_{\mathrm{pl}}(A_L) = \frac{1}{2} K_{\mathrm{eff},L} {A_L}^2 d - P_\parallel E d \left( \frac{2}{\pi} + \frac{dg\lambda}{2w_\parallel + dg} \right) A_L - \frac{d^2 P_\parallel^2}{2(2w_\parallel + dg)} E^2, \quad \text{(S17)}$$

where

$$K_{\mathrm{eff},L} = \frac{1}{2} K_0 + \frac{2 g w_\parallel \lambda^2}{2w_\parallel + dg}. \quad \text{(S18)}$$

The first term in Eq. (S17) is the intrinsic bulk restoring force originating from orientational-order distortion, whereas the second term arises from the coupling between the bulk director displacement and the interfacial anchoring.

The equilibrium bulk amplitude is obtained from

$$\frac{\partial F_{\mathrm{pl}}}{\partial A_L} = K_{\mathrm{eff},L} A_L d - P_\parallel E d \left( \frac{2}{\pi} + \frac{dg\lambda}{2w_\parallel + dg} \right) = 0. \quad \text{(S19)}$$

This gives

$$A_L = \frac{P_\parallel E \left( \frac{2}{\pi} + \frac{dg\lambda}{2w_\parallel + dg} \right)}{K_{\mathrm{eff},L}}. \quad \text{(S20)}$$

The polarization response should be evaluated from the spatial average of the induced polarization displacement. Within the present effective model,

$$\langle \delta P_L \rangle = \frac{1}{d} \int_0^d P_\parallel \left( \phi_A(z) + \phi_s \right) dz. \quad \text{(S21)}$$

Using Eq. (S1), this becomes

$$\langle \delta P_L \rangle = P_\parallel \left( \frac{2}{\pi} A_L + \phi_s \right). \quad \text{(S22)}$$

Substituting Eqs. (S16) and (S20), we obtain

$$\langle \delta P_L \rangle = P_\parallel^2 E \left[ \frac{\left( \frac{2}{\pi} + \frac{dg\lambda}{2w_\parallel + dg} \right)^2}{K_{\mathrm{eff},L}} + \frac{d}{2w_\parallel + dg} \right]. \quad \text{(S23)}$$

Therefore, the polarization susceptibility is

$$\chi_L = \frac{\partial \langle \delta P_L \rangle}{\partial E} = P_\parallel^2 \left[ \frac{\left( \frac{2}{\pi} + \frac{dg\lambda}{2w_\parallel + dg} \right)^2}{K_{\mathrm{eff},L}} + \frac{d}{2w_\parallel + dg} \right]. \quad \text{(S24)}$$

Thus, the dielectric strength is given by

$$\Delta\varepsilon_L \sim \frac{\chi_L}{\varepsilon_0} = \frac{P_\parallel^2}{\varepsilon_0} \left[ \frac{\left( \frac{2}{\pi} + \frac{dg\lambda}{2w_\parallel + dg} \right)^2}{K_{\mathrm{eff},L}} + \frac{d}{2w_\parallel + dg} \right]. \quad \text{(S25)}$$

The relaxation dynamics of the low-frequency mode are described by the overdamped equation of motion for $A_L$,

$$\gamma_L \frac{dA_L}{dt} = -K_{\mathrm{eff},L} A_L + P_{\parallel} E(t) \left( \frac{2}{\pi} + \frac{dg\lambda}{2w_{\parallel} + dg} \right), \tag{S26}$$

where $\gamma_L$ is the effective rotational viscosity. Therefore,

$$\tau_L = \frac{\gamma_L}{K_{\mathrm{eff},L}}, \tag{S27}$$

and

$$f_L = \frac{1}{2\pi\tau_L} = \frac{K_{\mathrm{eff},L}}{2\pi\gamma_L}. \tag{S28}$$

Equations (S25) and (S28) show how the planar low-frequency response depends on the balance between the intrinsic bulk restoring force and the interfacial contribution.

A particularly important limiting case is obtained when the intrinsic bulk restoring force dominates over the interfacial contribution,

$$\frac{1}{2} K_0 \gg \frac{2 g w_{\parallel} \lambda^2}{2w_{\parallel} + dg}. \tag{S29}$$

In this limit,

$$K_{\mathrm{eff},L} \simeq \frac{1}{2} K_0, \tag{S30}$$

and hence

$$f_L \simeq \frac{K_0}{4\pi\gamma_L}, \tag{S31}$$

which is nearly independent of the cell thickness. If, in addition, the interfacial polarization contribution dominates the dielectric amplitude and $2w_{\parallel} \gg dg$, Eq. (S25) reduces approximately to

$$\Delta\varepsilon_L \sim \frac{P_{\parallel}^2}{2\varepsilon_0 w_{\parallel}} d. \tag{S32}$$

Thus, the model explains the experimental trend in which the dielectric strength increases with cell thickness, whereas the relaxation frequency remains nearly thickness-independent. Physically, the relaxation frequency is governed mainly by the intrinsic bulk restoring force associated with orientational-order distortion, while the dielectric amplitude is enhanced by the polarization displacement accumulated across the cell and by the interfacial director response.

**I-2: Homeotropic director configuration under dc electric field**

We next consider the case in which a sufficiently strong dc electric field reorients the director toward the field direction. In this configuration, the low-frequency mode no longer requires a spatially distributed director deformation constrained by the substrates. Instead, the response is dominated by a local change in the orientational order parameter itself.

The order parameter is again defined as

$$\eta = \langle \cos^2 \theta \rangle. \tag{S33}$$

In the homeotropic state, the director is approximately parallel to the electric field, and the field-induced change in $\eta$ directly changes the polarization component along the field direction. We define

$$\delta\eta = \eta - \eta_0. \tag{S34}$$

The free energy for this order-parameter fluctuation is written as

$$F_{\text{homeo}}(\delta\eta) = \frac{1}{2} K_\eta \delta\eta^2 d - P_\eta E d \delta\eta, \tag{S35}$$

where $P_\eta$ is the effective coupling coefficient between the orientational-order change and the electric-field-induced polarization response.

The equilibrium value of $\delta\eta$ is obtained from

$$\frac{\partial F_{\text{homeo}}}{\partial \delta\eta} = 0, \tag{S36}$$

which gives

$$\delta\eta = \frac{P_\eta}{K_\eta} E. \tag{S37}$$

The dielectric susceptibility in the homeotropic state is therefore

$$\chi_L = \frac{\partial}{\partial E}\left(P_\eta \delta\eta\right) = \frac{P_\eta^2}{K_\eta}, \tag{S38}$$

and

$$\Delta\varepsilon_L \sim \frac{P_\eta^2}{\varepsilon_0 K_\eta}. \tag{S39}$$

The relaxation dynamics are described by

$$\gamma_L \frac{d\delta\eta}{dt} = -K_\eta \delta\eta + P_\eta E(t), \tag{S40}$$

where $\gamma_L$ is the effective viscosity for the orientational-order fluctuation. Thus,

$$\tau_L = \frac{\gamma_L}{K_\eta}, \tag{S41}$$

and

$$f_L = \frac{K_\eta}{2\pi\gamma_L}. \tag{S42}$$

Equations (S39) and (S42) show that, in the homeotropic configuration, the dielectric response directly reflects the intrinsic order-parameter fluctuation. Since the director deformation constrained by the substrates is largely removed, the response is no longer governed primarily by the interfacial anchoring term[1]. This explains why the dielectric anomaly associated with the low-frequency mode becomes more pronounced when the director is reoriented toward the electric-field direction. It should be noted, however, that the dc electric field plays two distinct roles. The first is to reorient the director and thereby remove the interfacially constrained director-deformation component. The second is to modify the fluctuation of the orientational distribution once the director is already aligned along the field. In this field-aligned state, increasing the dc bias can reduce the differential order-parameter response to the ac probing field (Fig. 3d). This reduction naturally leads to a decrease in the dielectric strength. After the low-frequency mode becomes pronounced, the subsequent decrease in relaxation frequency indicates that the field also lowers the effective dynamic response rate of the collective fluctuation, namely the ratio between the effective restoring coefficient and the effective kinetic coefficient (Fig. 3e). We therefore do not attribute the voltage dependence to a simple increase in restoring stiffness alone. Instead, the observed decrease in both dielectric strength and relaxation frequency suggests that the dc field suppresses the oscillator strength of the order-parameter fluctuation and slows the collective response in the field-aligned state.

Therefore, although the planar and homeotropic configurations differ in the extent to which the director-deformation component and interfacial anchoring contribute to the response, both cases originate from perturbations of the orientational order associated with molecular rotation around the short molecular axis. Thus, the low-frequency mode is expected to be sensitive to the softening of the orientational-order stiffness near the $SmZ_A$–$N_F$ phase transition. This provides a natural explanation for the pronounced dielectric anomaly and the soft-mode-like behavior observed for the low-frequency mode near the phase transition.

**Reference**

1. Nakase, M. *et al.* Field-induced phase transition behaviour of ferroelectric nematic liquid crystals under DC electric fields. *Soft Matter* **22**, 2538–2544 (2026).

**Supplementary Note II. Derivation of the high-frequency mode**

To understand the physical origin of the high-frequency dielectric relaxation mode, we consider a collective phase displacement of the transverse polarization component around the director. The molecular dipole is tilted with respect to the molecular long axis and therefore has a component perpendicular to the long axis. This transverse dipolar component can rotate around the long axis with a certain degree of freedom. Owing to dipole–dipole correlations, however, the azimuthal distribution of this transverse component is not completely isotropic, and a biased transverse polarization component remains at the macroscopic level. We denote this effective transverse polarization component as $P_{\perp}$. The high-frequency mode is then described as a collective displacement of this transverse polarization phase around the director. Importantly, this motion does not require an appreciable deformation of the director field itself.

We introduce a collective rotational displacement $\phi_A(z)$, which describes the phase displacement of the transverse polarization component $P_{\perp}$ along the cell normal direction $z$. Since the displacement is constrained by the interfaces and becomes largest in the bulk, we approximate the spatial profile by

$$\phi_A(z) = A_H \sin\left(\frac{\pi z}{d}\right), \tag{S43}$$

where $A_H$ is the maximum displacement amplitude at the cell center $z = d/2$, and $d$ is the cell thickness. The magnitude of the transverse polarization component is written phenomenologically as

$$P_{\perp} = P_s \sin\theta\,, \tag{S44}$$

where $P_s$ is the magnitude of the local polar order and $\theta$ represents the tilt angle of the molecular dipole from the director. Here, $P_{\perp}$ should be understood as the effective transverse polarization component that remains after azimuthal averaging under dipole–dipole correlations.

Although the director field remains essentially unchanged during this motion, the collective displacement of $P_{\perp}$ can have a weak bulk restoring energy. We write this polarization-displacement energy per unit area as

$$F_{\mathrm{pol}} = \int_0^d \frac{1}{2} K_P \phi_A(z)^2 dz, \tag{S45}$$

where $K_P$ is the bulk stiffness associated with the collective displacement of the transverse polarization phase. Substituting Eq. (S43), we obtain

$$F_{\mathrm{pol}} = \frac{1}{2} K_P A_H^2 \int_0^d \sin^2\left(\frac{\pi z}{d}\right) dz. \tag{S46}$$

Using

$$\int_0^d \sin^2\left(\frac{\pi z}{d}\right) dz = \frac{d}{2}, \tag{S47}$$

the free energy is

$$F_{\text{pol}} = \frac{1}{4} K_P A_H^2 d. \tag{S48}$$

We next introduce an interfacial displacement $\phi_s$, representing the phase displacement of the transverse polarization near the substrate surface. The interfacial contribution is written as

$$F_{\text{int}} = w_\perp \phi_s^2 + \frac{1}{2} dg(\phi_s - \lambda A_H)^2, \tag{S49}$$

where $w_\perp$ is the anchoring coefficient for the transverse polarization phase, $g$ is the coupling strength between the bulk and interfacial displacements, and $\lambda$ describes how efficiently the bulk displacement induces an interfacial displacement.

The electric field couples to the transverse polarization component through the collective displacement $\phi_A(z)$. The electric field energy is

$$F_E = -\int_0^d P_\perp E(\phi_A(z) + \phi_s) dz. \tag{S50}$$

Substituting Eq. (S43),

$$F_E = -P_\perp E \left( A_H \int_0^d \sin\left(\frac{\pi z}{d}\right) dz + \phi_s d \right). \tag{S51}$$

Using

$$\int_0^d \sin\left(\frac{\pi z}{d}\right) dz = \frac{2d}{\pi}, \tag{S52}$$

we obtain

$$F_E = -P_\perp E d \left(\frac{2}{\pi} A_H + \phi_s\right). \tag{S53}$$

Accordingly, the total free energy becomes

$$F_H(A_H, \phi_s) = \frac{1}{4} K_P A_H^2 d + w_\perp \phi_s^2 + \frac{1}{2} dg(\phi_s - \lambda A_H)^2 - P_\perp E d \left(\frac{2}{\pi} A_H + \phi_s\right). \tag{S54}$$

The equilibrium interfacial displacement is obtained from

$$\frac{\partial F_H}{\partial \phi_s} = 0. \tag{S55}$$

This gives

$$2 w_\perp \phi_s + dg(\phi_s - \lambda A_H) - P_\perp E d = 0. \tag{S56}$$

Therefore,

$$\phi_s = \frac{dg\lambda}{2w_\perp + dg} A_H + \frac{dP_\perp}{2w_\perp + dg} E. \tag{S57}$$

Substituting Eq. (S57) into Eq. (S54), we obtain an effective free energy for the collective amplitude,

$$F_H(A_H) = \frac{1}{2} K_{\text{eff},H} A_H^2 d - P_\perp E d A_H \left(\frac{2}{\pi} + \frac{dg\lambda}{2w_\perp + dg}\right) - \frac{d^2 P_\perp^2}{2(2w_\perp + dg)} E^2, \tag{S58}$$

with

$$K_{\text{eff},H} = \frac{1}{2}K_P + \frac{2gw_\perp\lambda^2}{2w_\perp + dg}. \tag{S59}$$

The first term represents the weak bulk stiffness of the collective transverse-polarization displacement, whereas the second term represents the interfacial restoring contribution. The equilibrium amplitude is obtained from

$$\frac{\partial F_H}{\partial A_H} = K_{\text{eff},H}A_H d - P_\perp E d\left(\frac{2}{\pi} + \frac{dg\lambda}{2w_\perp + dg}\right) = 0, \tag{S60}$$

giving

$$A_H = \frac{P_\perp E}{K_{\text{eff},H}}\left(\frac{2}{\pi} + \frac{dg\lambda}{2w_\perp + dg}\right). \tag{S61}$$

The average polarization response is evaluated from the spatial average of the induced displacement of the transverse polarization component. In the present effective description, both the bulk phase displacement $\phi_A(z)$ and the interfacial phase displacement $\phi_s$ contribute to the cell-averaged polarization response. Therefore,

$$\langle \delta P_H \rangle = \frac{1}{d}\int_0^d P_\perp \left(\phi_H(z) + \phi_s\right) dz. \tag{S62}$$

Using Eq. (S43), this becomes

$$\langle \delta P_H \rangle = P_\perp\left(\frac{2}{\pi}A_H + \phi_s\right). \tag{S63}$$

Substituting Eq. (S57),

$$\langle \delta P_H \rangle = P_\perp\left[\left(\frac{2}{\pi} + \frac{dg\lambda}{2w_\perp + dg}\right)A_H + \frac{dP_\perp}{2w_\perp + dg}E\right]. \tag{S64}$$

Using Eq. (S61), we obtain

$$\langle \delta P_H \rangle = P_\perp^2 E\left[\frac{\left(\frac{2}{\pi} + \frac{dg\lambda}{2w_\perp + dg}\right)^2}{K_{\text{eff},H}} + \frac{d}{2w_\perp + dg}\right]. \tag{S65}$$

Therefore, the polarization susceptibility is

$$\chi_H = \frac{\partial\langle \delta P_H \rangle}{\partial E} = P_\perp^2\left[\frac{\left(\frac{2}{\pi} + \frac{dg\lambda}{2w_\perp + dg}\right)^2}{K_{\text{eff},H}} + \frac{d}{2w_\perp + dg}\right]. \tag{S66}$$

Thus, the dielectric strength is given by

$$\Delta\varepsilon_H \sim \frac{\chi_H}{\varepsilon_0} = \frac{P_\perp^2}{\varepsilon_0}\left[\frac{\left(\frac{2}{\pi} + \frac{dg\lambda}{2w_\perp + dg}\right)^2}{K_{\text{eff},H}} + \frac{d}{2w_\perp + dg}\right]. \tag{S67}$$

In the experimentally relevant regime, the bulk polarization-phase stiffness is assumed to be much smaller than the interfacial restoring contribution,

$$\frac{1}{2}K_P \ll \frac{2gw_\perp\lambda^2}{2w_\perp + dg}, \tag{S68}$$

and, for sufficiently thick cells,

$$dg \gg 2w_\perp. \tag{S69}$$

Under this condition, the effective restoring coefficient becomes

$$K_{\mathrm{eff},H} \simeq \frac{2w_\perp\lambda^2}{d}. \tag{S70}$$

If the amplitude-mediated contribution dominates the dielectric response, i.e.,

$$\frac{\left(\frac{2}{\pi} + \frac{dg\lambda}{2w_\perp + dg}\right)^2}{K_{\mathrm{eff},H}} \gg \frac{d}{2w_\perp + dg} \tag{S71}$$

we obtain

$$\Delta\varepsilon_H \sim \frac{P_\perp^2 d}{2\varepsilon_0 w_\perp \lambda^2}\left(\frac{2}{\pi} + \lambda\right)^2. \tag{S72}$$

The dynamics of the collective amplitude are described by the overdamped equation

$$\gamma_H \frac{dA_H}{dt} = -K_{\mathrm{eff},H}A_H + P_\perp E(t)\left(\frac{2}{\pi} + \frac{dg\lambda}{2w_\perp + dg}\right), \tag{S73}$$

where $\gamma_H$ is the effective viscosity for the collective transverse-polarization phase displacement. The relaxation time is

$$\tau_H = \frac{\gamma_H}{K_{\mathrm{eff},H}}, \tag{S74}$$

and the relaxation frequency is

$$f_H = \frac{1}{2\pi\tau_H} = \frac{K_{\mathrm{eff},H}}{2\pi\gamma_H}. \tag{S75}$$

Using Eq. (S70),

$$f_H \simeq \frac{w_\perp\lambda^2}{\pi\gamma_H}\frac{1}{d}. \tag{S76}$$

Combining Eqs. (S72) and (S76), we obtain

$$\Delta\varepsilon_H f_H \simeq \frac{P_\perp^2}{2\pi\varepsilon_0\gamma_H}\left(\frac{2}{\pi} + \lambda\right)^2, \tag{S77}$$

which is independent of cell thickness and anchoring strength. Therefore, the present model naturally explains the experimentally observed relations, $\Delta\varepsilon_H \propto d$, $f_H \propto d^{-1}$, and $\Delta\varepsilon_H f_H \approx \mathrm{const.}$

The high-frequency mode is thus assigned to a collective phase displacement of the transverse polarization component around the director. This motion does not involve appreciable director deformation, and therefore the Frank elastic contribution is negligible. Although a weak bulk stiffness associated with the collective polarization displacement can exist, the observed thickness scaling is reproduced when this bulk stiffness is much smaller than the boundary-induced restoring contribution. In this sense, the mode is Goldstone-like: the bulk transverse-polarization phase is nearly degenerate, and a finite relaxation frequency appears mainly through boundary-induced symmetry breaking.